\documentclass[twocolumn,prl,superscriptaddress]{revtex4}
\usepackage[colorlinks,linkcolor=blue,anchorcolor=blue,citecolor=blue,filecolor=blue,menucolor=blue,runcolor=blue,urlcolor=blue,frenchlinks=blue]{hyperref}
\usepackage{amsmath}
\usepackage{graphicx}
\usepackage{amssymb}
\usepackage{dcolumn}
\usepackage{mathrsfs}
\usepackage{bm}
\begin{document}

\title{{Measurement of spin Chern numbers in quantum simulated topological insulators}}

\author{Qing-Xian Lv}
\affiliation{Guangdong Provincial Key Laboratory of Quantum Engineering and Quantum Materials, School of Physics and Telecommunication Engineering, South China Normal University, Guangzhou 510006, China}

\affiliation{Guangdong-Hong Kong Joint Laboratory of Quantum Matter, Frontier Research Institute for Physics, South China Normal University, Guangzhou 510006,
China}

\author{Yan-Xiong Du}
\email{yanxiongdu@m.scnu.edu.cn}
\affiliation{Guangdong Provincial Key Laboratory of Quantum Engineering and Quantum Materials, School of Physics and Telecommunication Engineering, South China Normal University, Guangzhou 510006, China}

\author{Zhen-Tao Liang}
\affiliation{Guangdong Provincial Key Laboratory of Quantum Engineering and Quantum Materials, School of Physics and Telecommunication Engineering, South China Normal University, Guangzhou 510006, China}

\author{Hong-Zhi Liu}
\affiliation{Guangdong Provincial Key Laboratory of Quantum Engineering and Quantum Materials, School of Physics and Telecommunication Engineering, South China Normal University, Guangzhou 510006, China}

\author{Jia-Hao Liang}
\affiliation{Guangdong Provincial Key Laboratory of Quantum Engineering and Quantum Materials, School of Physics and Telecommunication Engineering, South China Normal University, Guangzhou 510006, China}

\author{Lin-Qing Chen}
\affiliation{Guangdong Provincial Key Laboratory of Quantum Engineering and Quantum Materials, School of Physics and Telecommunication Engineering, South China Normal University, Guangzhou 510006, China}

\author{Li-Ming Zhou}
\affiliation{Guangdong Provincial Key Laboratory of Quantum Engineering and Quantum Materials, School of Physics and Telecommunication Engineering, South China Normal University, Guangzhou 510006, China}

\author{Shan-Chao Zhang}
\affiliation{Guangdong Provincial Key Laboratory of Quantum Engineering and Quantum Materials, School of Physics and Telecommunication Engineering, South China Normal University, Guangzhou 510006, China}

%\affiliation{ Guangdong Provincial Key Laboratory of Nuclear Science, Institute of Quantum Matter, South China Normal University, Guangzhou 510006, China}
\affiliation{Guangdong-Hong Kong Joint Laboratory of Quantum Matter, Frontier Research Institute for Physics, South China Normal University, Guangzhou 510006,
China}

\author{Dan-Wei Zhang}
\affiliation{Guangdong Provincial Key Laboratory of Quantum Engineering and Quantum Materials, School of Physics and Telecommunication Engineering, South China Normal University, Guangzhou 510006, China}
\affiliation{Guangdong-Hong Kong Joint Laboratory of Quantum Matter, Frontier Research Institute for Physics, South China Normal University, Guangzhou 510006,
China}

%\author{Tao Zhou}
%\affiliation{Guangdong Provincial Key Laboratory of Quantum Engineering and Quantum Materials, School of Physics and Telecommunication Engineering, South China %Normal University, Guangzhou 510006, China}

%\affiliation{Guangdong-Hong Kong Joint Laboratory of Quantum Matter, Frontier Research Institute for Physics, South China Normal University, Guangzhou 510006,
%China}

\author{Bao-Quan Ai}
\affiliation{Guangdong Provincial Key Laboratory of Quantum Engineering and Quantum Materials, School of Physics and Telecommunication Engineering, South China Normal University, Guangzhou 510006, China}

\affiliation{Guangdong-Hong Kong Joint Laboratory of Quantum Matter, Frontier Research Institute for Physics, South China Normal University, Guangzhou 510006,
China}

\author{Hui Yan}
\email{yanhui@scnu.edu.cn}
\affiliation{Guangdong Provincial Key Laboratory of Quantum Engineering and Quantum Materials, School of Physics and Telecommunication Engineering, South China Normal University, Guangzhou 510006, China}

\affiliation{Guangdong-Hong Kong Joint Laboratory of Quantum Matter, Frontier Research Institute for Physics, South China Normal University, Guangzhou 510006,
China}

\author{Shi-Liang Zhu}
\email{slzhu@scnu.edu.cn}
\affiliation{Guangdong Provincial Key Laboratory of Quantum Engineering and Quantum Materials, School of Physics and Telecommunication Engineering, South China Normal University, Guangzhou 510006, China}

\affiliation{Guangdong-Hong Kong Joint Laboratory of Quantum Matter, Frontier Research Institute for Physics, South China Normal University, Guangzhou 510006,
China}

%\date{\today}

\begin{abstract}
The topology of quantum systems has become a topic of great interest since the discovery of topological insulators. However, as a hallmark of the topological insulators, the spin Chern number has not yet been experimentally detected. The challenge to directly measure this topological invariant lies in the fact that this spin Chern number is defined based on artificially constructed wavefunctions. Here we experimentally mimic the celebrated Bernevig-Hughes-Zhang model with cold atoms, and then measure the spin Chern number with the linear response theory. We observe that, although the Chern number for each spin component is ill defined, the spin Chern number measured by their difference is still well defined when both energy and spin gaps are non-vanished.
%Our results may trigger the detection of all kinds of topological invariants in condensed matter physics and artificial quantum systems.
\end{abstract}

 \maketitle

{\sl Introduction.--} Topological matter refers to systems in which topology is required for their characterisation. Typical examples include matter with the quantum Hall effect, topological insulators, and  Dirac/Weyl semimetals \cite{Hasan2010,XQi2011,DWZhang2018,Ozawa2019}.
These systems are classified by certain robust topological
invariants. For instance, the quantum Hall effect is characterized
by the Chern number (CN) \cite{Thouless1982,Thouless1983}, which quantifies the number of the chiral edge states and is the fundamental reason  for the stability of the quantum Hall effect.

 In 1988, Haldane constructed a famous topological insulator model with the quantum Hall effect without Landau levels \cite{Haldane1988}. More recently, Kane and Mele proposed that a graphene should be a $Z_2$ topological
insulator with the  quantum spin Hall phase  \cite{Kane2005,Kane2005Z2}. This phase is a time reversal invariant state with a bulk band gap that supports the transport of spin in gapless edge states.
Analogous to the CN classification of the quantum Hall effect, a topological invariant called spin Chern number (SCN) was proposed by Haldane group \cite{DNSheng2006,LSheng2005} to address the stability of the quantum spin Hall effect. This arouse extensive interest and attention among numerous researches %This created broad interests among numerous researches
\cite{Hasan2010,XQi2011,DWZhang2018,Ozawa2019,LSheng2013,Prodan2009,LFu2006,Fukui2007,SLZhu2013,Carpentier2015,Canonico2019,CZChen2019,LSheng2014} since the SCN can characterise the helical edge states and explain the stability of the quantum spin Hall state under broken time-reversal symmetry \cite{LSheng2013} and its robustness against disorders \cite{DNSheng2006}. However, spin-orbit coupling is too weak to create a quantum spin Halll effect in graphene. Bernevig, Hughes, and Zhang (BHZ) proposed an experimentally realizable  model \cite{Bernevig2006}, which becomes one of the famous models for topological insulator research \cite{Hasan2010,XQi2011,DWZhang2018}. The BHZ  model has been realized in real condensed matter systems, where the theoretical predictions of quantized spin Hall conductance and metallic surface state are observed \cite{Konig2007,Hsieh2008,Hsieh2009}. However, measuring the associated topological invariants, the hallmarks of topological insulators with time-reversal symmetry, hasn't been realized.

Although topological invariants play a fundamental role in topological matter,   only CNs related to the quantum Hall
effect have been measured \cite{Sugawa2018,XTan2018,MYu2020,XTan2019,Duca2015,Roushan2014,TLi2016,Aidelsburger2015,Schweizer2016,Lohse2018}. The other key topological invariants, such as the $Z_2$ topological invariant and SCN, have not been experimentally detected since both of them are defined based on the artificially constructed wavefunctions which are difficult to realize in real condensed matter systems.

Here we report our quantum simulation of the celebrated BHZ Hamiltonian with cold atomic gas and the measurement of  the spin Berry curvature as well as the SCN.  We carefully designed a four-level atomic quantum gas to simulate the four-band  BHZ model with two pseudospins. Using  this well-controlled quantum system, we can independently create and manipulate each pseudospin's wavefunctions, which has not been done before and is necessary for measuring SCNs.  To extract the SCNs, we developed a method to evaluate the local Berry curvature for each pseudospin through nonadiabatic responses of the system. When the intermediate coupling between two pseudospins is absent, we observed that the CN for each spin component is well defined and the SCN is the difference between the CNs for the two pseudospins. Remarkably, in the presence of intermediate coupling, although the CN for each spin component is ill defined since the pseudospins are non-conserved, the SCN itself is well defined when both energy and spin gaps are non-vanished.
 \\

%%%%%%%%%%%%%%%%%%%%%%%%%%%%%%%%%%%%%%%%%%%%%%%%%
%%%%%%%%%%%%%%%%%%%%%%%%%%%%%%%%%%%%%%%%%%%%%%%%%%%
\begin{figure}[ptb]
\begin{center}
\includegraphics[width=8.0cm]{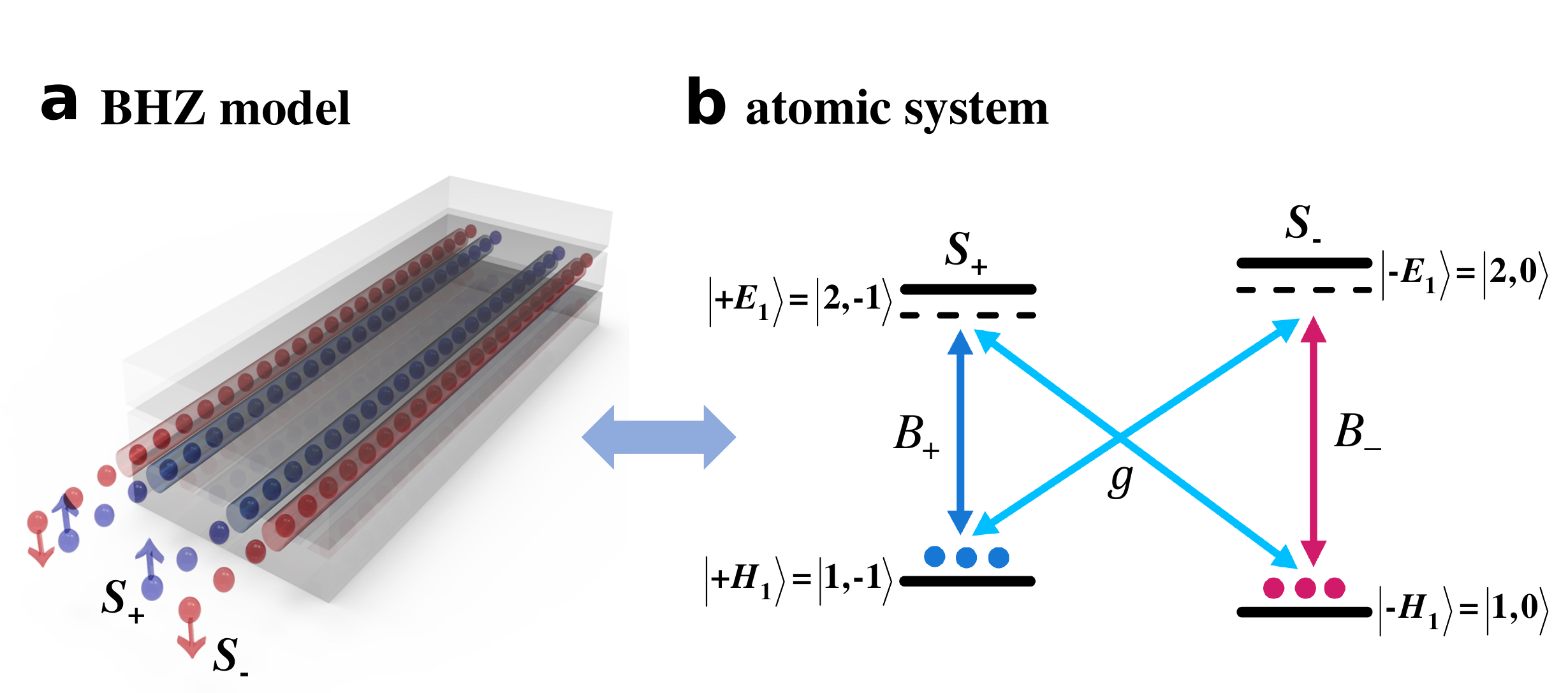}
\caption{\label{fig:set up} Mapping the BHZ model into an atomic system. (a) The BHZ model is a four-band tight-binding model in the basis $\{S_{+}, S_{-}\}$, where { the block $S_{+}=\{|+ E_1\rangle, |+ H_1\rangle \}$} is for the up-spin charge
carriers and the block {  $S_{-}=\{|- E_1\rangle, |- H_1\rangle \}$} is for the down-spin charge
carriers.  The model exhibits a quantum spin Hall effect characterized by a helical edge state: the carriers with opposite
spins move in opposite directions on a given edge. (b) Mapping the BHZ Hamiltonian into a four-level atomic system. In a cold $^{87}$Rb atomic system, we encode the basis of the BHZ model with four atomic levels as follows: $|+E_1\rangle=|F=2,m_F=-1\rangle$, $|+H_1\rangle=|F=1,m_F=-1\rangle$, $|-E_1\rangle=|F=2,m_F=0\rangle$ and $|-H_1\rangle=|F=1,m_F=0\rangle$. The coefficients $\mathbf{B}_{+}$ and $\mathbf{B}_{-}$ in Eqs. (\ref{BHZ})  can be independently manipulated with $\pi$-transition microwaves, and the coupling term $g$ between $\{|+E_1\rangle,|-H_1\rangle\}$ $(\{|-E_1\rangle,|+H_1\rangle\})$ can be manipulated with  $\sigma^{-}$ ($\sigma^{+}$)-transition microwaves. }
\end{center}
\end{figure}

%%%%%%%%%%%%%%%%%%%%%%%%%%%%%%%%%%%%%%%%%%%%%%%%%
%%%%%%%%%%%%%%%%%%%%%%%%%%%%%%%%%%%%%%%%%%%%%%%%%%%
\begin{figure}[ptb]
\begin{center}
\includegraphics[width=8.5cm]{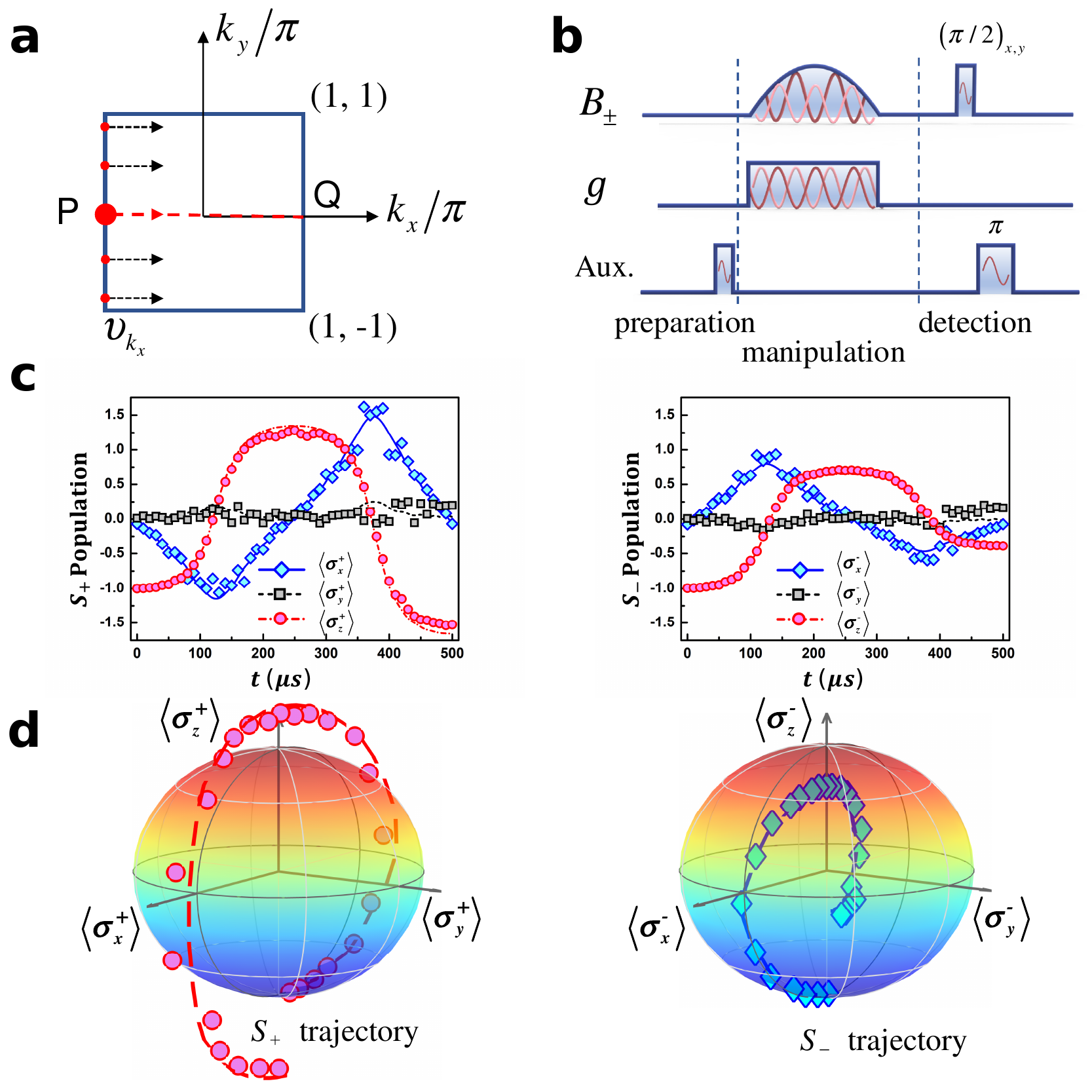}
\caption{\label{fig:evolution} Detection method using the linear response theory and tomographic results. (a) The procedure of measuring the Berry curvature in the first Brillouin zone. Manipulation based on the linear response theory is realized by sweeping $k_x$ (along the black arrows) at each $k_y$ with a ratio $\upsilon_{k_x}$. (b) Pulse sequences in the experiments. The system is initially prepared with an equal superposition of the  two lowest eigenstates. The effective Hamiltonian with the coefficients $\mathbf{B}_\pm$ can be realized with the control microwaves with  time-varying amplitudes, frequencies, and phases. Intermediate couplings $g$ can be used to mimic the non-conversing spin current case. In the detection step, we measure the population of each pseudospin to extract the Berry curvature. Auxiliary pulses (labelled by Aux.) are used for preparation and detection. (c) Tomography of pseudospin $S_\pm$ which corresponds to  parameter quenches along the line denoted by points $P$ and $Q$ in (a). Dashed and solid lines are  theoretical simulations with the Sch\"{o}dinger equation using  Hamiltonian (1). The evolving four-level state is divided into two pseudospins $S_\pm$. Diamonds, squares, and circles are  experimental data. (d) Trajectories of pseudospins $S_\pm$ in (c). Dashed lines are  theoretical simulations, and diamonds and circles are experimental data. Parameters: $M=2B $ and {$g=0.15A$}.
}
\end{center}
\end{figure}

%%%%%%%%%%%%%%%%%%%%%%%%%%%%%%%%%%%%%%%%%%%%%%%%%%%%%%%%
{\sl The BHZ model.--} In a pioneer paper \cite{Bernevig2006},
 Bernevig, Hughes, and Zhang proposed an effective model of the two-dimensional time-reversal-invariant topological insulator in HgTe/CdTe quantum wells as follows,
 %%%%%%%%%%%%%%%%%%%%%%%%%%%%%%%%%
 %%%%%%%%%%%%%%%%%%%%%%%%%%%%%%%
 \begin{equation}
\label{Hg} H_{\text{BHZ}}=\left(\begin{array}{cc} H_{+} & g\sigma_x \\ g\sigma_x & H_{-}\end{array}\right),
\end{equation}
 %%%%%%%%%%%%%%%%%%%%%%%%%%%%%%%%%
 %%%%%%%%%%%%%%%%%%%%%%%%%%%%%%%%%%%%%%%%%
%\begin{equation}
%\label{Hg} H_{\text{BHZ}}=\mathbf{B}_{+}\vec{\sigma}_{+}\otimes I_{-}+I_{+}\otimes\mathbf{B}_{-}\vec{\sigma}_{-}
%+g \sigma^x_{+}\otimes\sigma^x_{-},
%\end{equation}
where $H_\tau=\mathbf{B}_{\tau} \cdot \vec{\sigma}$  ($\tau=+,-)$ with   $\vec{\sigma}$  being the Pauli matrixes.  The coefficients $\mathbf{B}_\tau$ are given by
\begin{equation}
\label{BHZ} \left\{\begin{split}
\mathbf{B}_{+} =\mathbf{B}=(A\sin k_x,-A\sin k_y,M(\mathbf{k})),\\
 \mathbf{B}_{-}=(-A\sin k_x,-A\sin k_y,M(\mathbf{k})),\\
 M(\mathbf{k}) = M-2B(2-\cos k_x-\cos k_y).
 \end{split}\right.
\end{equation}
Here $A, M$, and $B$ are material parameters  dependent on the quantum well geometry and the momentum ${k_x, k_y} \in [-\pi, \pi]$.
%The basis in Eq. (\ref{Hg}) is  four subbands in the system and is ordered as $\{|+,E_1\rangle,|+,H_1\rangle,|-,E_1\rangle,|-,H_1\rangle\}$.
The $g$ describes the band coupling strength and this coupling term  between blocks is determined by crystal symmetry and plays an important role in determining the spin orientation of the helical edge state~\cite{XQi2011}. The basis in Eq. (\ref{Hg}) is  four subbands in the system and is ordered as $\{|+,E_1\rangle,|+,H_1\rangle,|-,E_1\rangle,|-,H_1\rangle\}$.
%Here we choose  $\Pi_{+-}=\sigma_x\otimes\sigma_x$ without the loss of generality.

%When $g=0$, $H_\text{BHZ}$ consists of two decoupled blocks that are topologically equivalent to two copies of the Haldane model \cite{Haldane1988}.

The topological properties of the BHZ model are well characterized by  a SCN \cite{LSheng2005}. The physics can be easily understood when the coupling strength $g=0$. Under this condition, each pseudospin  is
conserved, and thus $H_\text{BHZ}$ consists of two decoupled blocks that are topologically equivalent to two copies of the Haldane model \cite{Haldane1988}.
% where pseudospin $\tau$ is conserved.
Each
pseudospin has an independent CN $C_+$ and $C_{-}$ defined as $C_\tau=\frac{1}{2\pi}\int_{BZ}d\mathbf{k}\cdot\mathbf{\Omega_\tau}(\mathbf{k})$, where $\mathbf{\Omega_\tau}(\mathbf{k})=\mathbf{B}_{\tau}/|\mathbf{B}_{\tau}|$ is the Berry curvature in the first Brillouin zone (BZ).
The time-reversal symmetry gives rise to a vanished total CN ( i.e., $C_{+}+C_{-}=0$). The
difference
\begin{equation}
C_s=(C_+-C_-)/2
\end{equation}
is 1 for $M/2B>0$ and defines  quantized spin Hall conductivity, and it is 0 for $M/2B<0$. %The SCN gives the same topological classification as the $Z_2$ invariant introduced by Kane and Mele in Ref. \cite{Kane2005}.

A fundamental result of the BHZ model is that  the SCN can still be well defined  when $\langle\sigma_z\rangle$ is non-conserved under the condition that two blocks are coupled.  The energy gap of the system is defined as $\Delta_{cv}\equiv \mathrm{Min}(\mathcal{E}_3-\mathcal{E}_1)=\mathrm{Min}|2\mathcal{E}_1|$, where
the eigenvalues of the Hamiltonian (\ref{Hg}) are obtained as $\mathcal{E}_1=\mathcal{E}_2=-\sqrt{B_x^2+B_y^2+B_z^2+g^2}$ and $\mathcal{E}_3=\mathcal{E}_4=-\mathcal{E}_1$.  To calculate the spin spectrum gap, we first project the system to the subsystem spanned by the two lowest eigenstates $|\phi_{1,2}\rangle$ of $\mathcal{E}_{1,2}$, which induces a reduced Hamiltonian $H^s$ with $H^s_{jl}=\langle\phi_j|I\otimes \sigma_z|\phi_l\rangle, (j,l=1,2)$ with $I$ being the identity matrix. Through diagonalizing $H^s$ to obtain the eigenvalues $\mathcal{E}_\pm$ and the related eigenstates $|\psi_{\pm}\rangle$, we can define the spin spectrum gap $\Delta_s=|\mathcal{E}_+-\mathcal{E}_-|$ \cite{Prodan2009,LSheng2014}. Initially, it is found that the SCN may change its sign when the Hamiltonian is continuously deformed  using spin rotation while the bulk gap is kept unchanged \cite{LFu2006,Fukui2007}. However, it has been shown that the SCN is well defined if both the energy gap and the spin spectrum gap are non-vanishing \cite{Prodan2009}. One can define a gauge potential $A_{k_\mu}^\tau \equiv \langle\psi_\tau|\partial _{k_\mu} |\psi_\tau\rangle$ and the related gauge field $F^\tau_{k_x k_y} \equiv \partial_{k_x} A^\tau_{k_y}-\partial_{k_y} A^\tau_{k_x}$.  Integrating the spin Berry curvature defined by $F_{k_xk_y}^s=F_{k_xk_y}^{+}-F_{k_xk_y}^{-}$  over the BZ  gives the SCN of the system
%%%%%%%%%%%%%%%%%%%%%%%%%%
\begin{equation}
\label{Cs2} C_s=\frac{1}{2}\int_{\text{BZ}}
F_{k_xk_y}^s dk_xdk_y
\end{equation}
under the conditions $\Delta_s>0$ and  $\Delta_{cv}>0$. For any $g$ in the BHZ model, the SCN is 1 for $M/2B>0$ and  is 0 for $M/2B<0$, while $M/2B=0$ is a critical point of the topological phase transition.  Because the eigenfunctions $|\psi_{\pm}\rangle$  are relevant to  an artificially-constructed Hamiltonian $H^s$, not to the original Hamiltonian of the system, it is difficult to directly use definition (\ref{Cs2}) to  measure the SCNs in a real condensed matter system, and thus the SCNs of a quantum system have  not yet been experimentally observed. However, this problem can be solved in  well-designed artificial quantum systems where  wavefunctions of the reduced Hamiltonian $H^s$ can be created and manipulated.   \\

{\sl Quantum simulation of the BHZ model with cold atoms.--} As shown in Fig. 1 and deduced in detail in Supplemental Material (SM) \cite{SM}, the BHZ model can be mapped into a four-level $^{87}$Rb atomic system with the codes: $S_+=\{|+E_1\rangle=|F=2,m_F=-1\rangle, |+H_1\rangle=|F=1,m_F=-1\rangle\}$, $S_{-}=\{|-E_1\rangle=|F=2,m_F=0\rangle, |-H_1\rangle=|F=1,m_F=0\rangle\}$. The coefficients $\mathbf{B}_\pm$ and the coupling strength $g$ can be realized with the microwaves coupling the atomic energy levels.
The $^{87}$Rb atoms are evaporatively cooled down and then trapped in an optical dipole trap with atom number about $10^6$ and temperature about 10 $\mu$K.  A magnetic field about 0.5 Gauss is applied along the $z$ direction as the quantization axis, which generates a $700$ kHz frequency difference between the two states' energy level difference in $S_{+}$ and that in $S_{-}$. Thus, the quantum state of pseudospins $S_{+}$ and $S_{-}$ can be manipulated independently using $\pi$-transition microwaves. The $g\Pi_{+-}$ term between the pseudospins  $\{|+E_1\rangle,|-H_1\rangle\}$ $(\{|-E_1\rangle,|+H_1\rangle\})$ can be manipulated with  $\sigma^{-}$ ($\sigma^{+}$)-transition microwaves.
%forming a $\infty$-type configuration. Since the $\sigma^{\pm}$-transition are synchronized in the same frequencies, amplitudes and phases, they can be %realized by a single microwave. Through tilting the orientation of a circular polarization horn antenna, we will achieve the same interaction strength of %$\sigma^\pm$ -transitions with the aid of the back reflection of the microwave, even though the transition coefficients of $\sigma^{\pm}$-transition are not %equal.
In our system, the $1/e$ coherence time  $T_2^*$ of $S_{+}$ ( magnetism-sensitive sublevel) is about 4 ms after stabilizing the quantization axis with active feedback control, while the coherence time  $T_2^*$ of $S_{-}$ (magnetism-insensitive sublevel) is longer than 1 second.
The system can operate under a maximal Rabi frequency of tens kHz \cite{Sugawa2018}, so it allows sufficiently fast manipulations and all manipulations can be finished within the decoherence times of  $S_+$ and $S_-$. \\
 %Since no entanglement is needed between the pseudospins, as well as the population flow between the ensemble-encoding pseusospins can mimic the inconserving spin flows, our fully-controllable system is naturally suitable for simulating the models of topological insulators.\\

%%%%%%%%%%%%%%%%%%%%%%%%%%%%%%%%%%%%%%%%%%%%%%%%%%%%%%%%
%%%%%%%%%%%%%%%%%%%%%%%%%%%%%%%%%%%%%%%%%%%%%%%%%%
\begin{figure*}[ptb]
\begin{center}
\label{Tomographic}
\includegraphics[width=13cm]{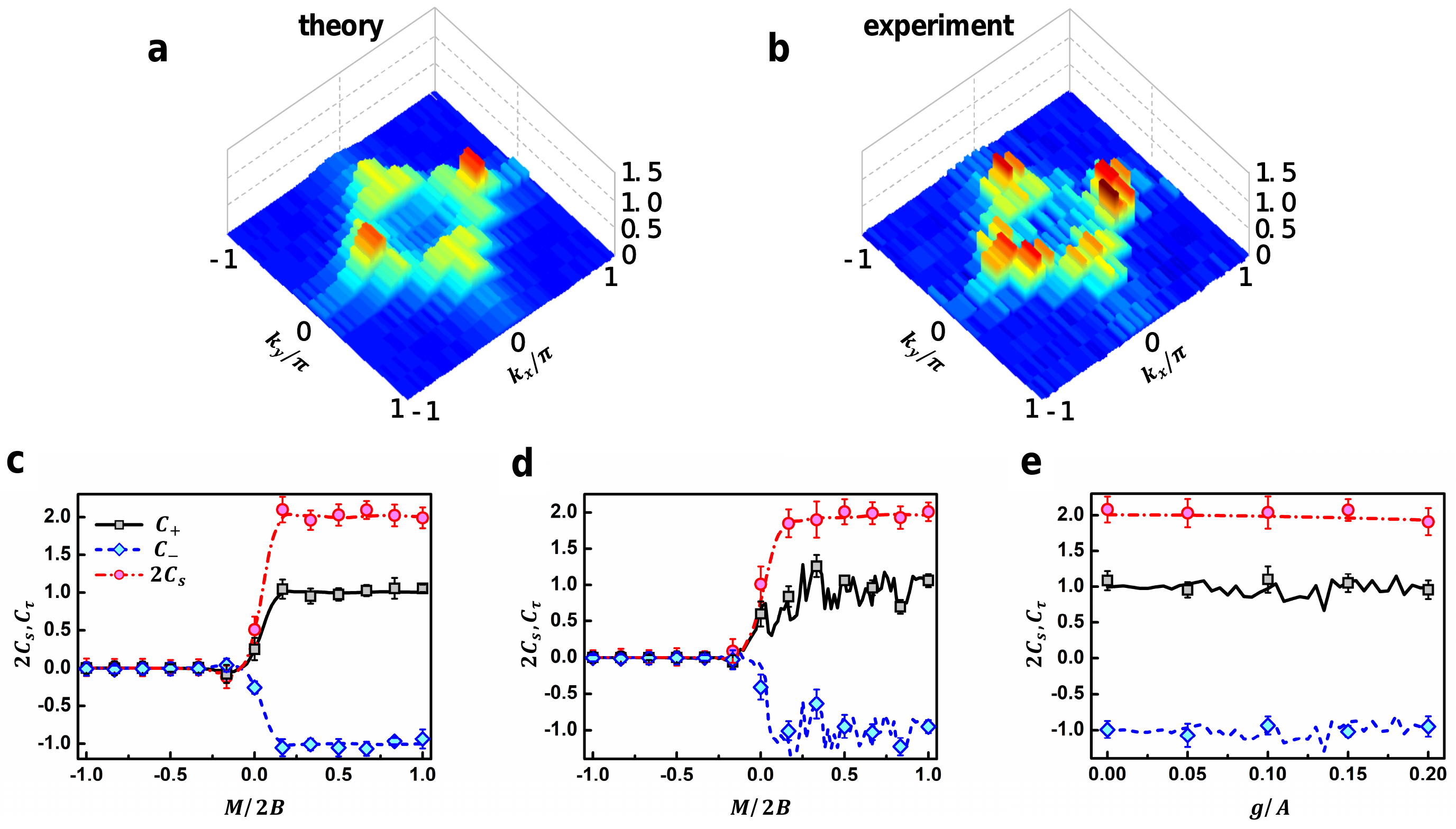}
\caption{\label{topological transition} Spin Berry curvatures and SCNs. (a)(b) Spin Berry curvatures in the first Brillouin zone. (a) Theoretical simulation. The spin Berry curvature is computed by the difference between the Berry curvature of each pseudospin using the linear response theory with Eq. (\ref{Fxy}). (b) Experimental measurement.  (c) SCNs $C_s$ and CN $C_{\pm}$ versus $M/2B$ with $g=0$. (d) $C_s$ and  $C_{\pm}$ versus $M/2B$ with $g=0.15A$.  A topological transition occurs at $M/2B=0$ in (c,d).  (e) $C_s$ and  $C_{\pm}$ against intermediate coupling $g$ at $M=2B$. Lines: theoretical curves  computed by the linear response theory with Eq. (\ref{Fxy}). Circles, squares, and diamonds: experimental data. Red dashed-dotted lines with red circles: $C_s$. Black solid lines with black squares: $C_+$. Blue dashed lines with blue diamonds: $C_-$. The SCNs have been multiplied by a factor of 2. The error bars are standard error of the 4 measurements.  }
\end{center}
\end{figure*}
%%%%%%%%%%%%%%%%%%%%%%%%%%%%%%%%%%%%%%%%%%%%%%%%%%%%%%%%

{\sl Measuring spin Berry curvature.--} We now show that in this simulated BHZ model, the spin Berry curvature defined in the first BZ $k_x, k_y \in [-\pi, \pi]$ can be detected using the linear response theory, which is given as the leading order correction to adiabatic
manipulation.

Our measuring procedure is described in Fig. 2.
%We sweep $k_x$ at each $k_y$ to cover the whole first Brillouin zone.
In the experiments, we choose one $k_y$ in the set $\{k_y (q)=-\pi+\frac{\pi}{5}(q-1), q=1,...,11\}$, and then sweep $k_x$ (along the black arrows in Fig. 2(a)) at each $k_y$ with a ratio $v_{k_x}=2\pi/T$, i.e.,  $k_x(t)=v_{k_x}t-\pi$ with the time $t\in [0,T]$. The experiment sequence consists of preparation, manipulation, and detection steps, as shown in Fig. 2(b). In the preparation step, we prepare an initial state given by { $|\Psi(0)\rangle=\left ( \begin{array}{c}
|\eta^+(0)\rangle\\
|\eta^-(0)\rangle
\end{array}\right)$, and $\left ( \begin{array}{c}
1\\
0
\end{array}\right)\otimes|\eta^+(0)\rangle=|\phi_1(0)\rangle, \left ( \begin{array}{c}
0\\
1
\end{array}\right)\otimes|\eta^-(0)\rangle=|\phi_2(0)\rangle$, where $|\eta^\tau(t)\rangle$ is the wave function defined in pseudospin $S_\tau$ } and  $|\phi_{1,2}(0)\rangle$ are the two lower normalized eigenstates of the Hamiltonian $H(k_x,k_y)$ at $k_x=-\pi$ and $k_y=k_y(q)$ for a specific $q$.   The whole wavefunction $|\Psi(0)\rangle$ is however normalized to $2$; that is, $\langle \Psi (t) |\Psi(t)\rangle=2$ since it is more convenient if the wavefunction of each pseudospin itself is normalized to one.

In the manipulation step, we sweep the Hamiltonian from $H(-\pi,k_y(q))$ to $H(k_x(t),k_y(q))$ by  modulating the amplitudes, phases, and frequencies of the microwaves used to manipulate the atoms. To realize the BHZ Hamiltonian, the relative phases between the microwaves should be carefully determined by the interference of microwave driven Rabi oscillations. To mimic the non-conversing spin currents, the intermediate coupling $g$ is introduced and can be controlled by the amplitude of the microwave coupling between the pseudospins.  To keep two-photon resonance condition with $B_\pm$, the frequencies of the intermediate coupling should also be swept.
% For instance, the generalized force on each pseudopsin can be obtained by
{ Following the method outlined in Ref.\cite{Gritsev2012},  we show in SM \cite{SM} that the Berry
curvature of pseudospin $S_{\tau}$ can be derived by }
\begin{equation}
\label{Fxy}
F_{k_xk_y}^\tau=\frac{\langle f_{k_y}^\tau\rangle-\langle f_0^\tau\rangle}{v_{k_x}},
\end{equation}
%%%%%%%%%%%%%%%%%%%%%%%%%%%%%%%%%%%%%%%%%%%%%%%%%%%%%%%%%%%%%
where $\langle f_{k_y}^\tau\rangle= -\langle \partial_{k_y} H_\tau
\rangle=A\cos k_y\langle \sigma_y^\tau\rangle+2B\sin k_y\langle \sigma_z^\tau\rangle$ {($\sigma^\tau_{x,y,z}$ are Pauli matrices for pseudospin $S_{\tau}$)} and $\langle f_0^\tau\rangle=4B\sin k_y$. Therefore, the Berry curvature can be derived by tomographically
measuring the expectation values $\langle\sigma^\tau_{y,z}\rangle$, which is described in detail in SM \cite{SM}. As emphasized in SM \cite{SM}, { the expectation values $\langle\sigma^\tau_{x,y,z}\rangle$ are defined based on the pseudospin wavefunctions $|\eta^\tau(t)\rangle$}.
% $|\eta^\tau (t)\rangle$.
%%%%%%%%%%%%%%%%%%%%%%%%

In our experiments, the parameters are set to be $A=B=2\pi\times24$ kHz and $T=500$ $\mu$s, which induce a nonadiabatic condition of $AT=BT=24\pi$, as discussed in SM \cite{SM}. The experimental data  $\langle\sigma_{x,y,z}^\tau\rangle$ are plotted in Fig. 2(c)  when $(k_x,k_y)$ is swept from $(k_x(t=0),k_y (q))=(-\pi,0)$ with the parameters $M=2B$ and {$g=0.15A$}. From these data, we plot the evolution trajectories $\langle\sigma_{x,y,z}^\tau\rangle$ of $S_\pm$, where the red (blue) dashed lines are theoretical curves and red circles (blue diamonds) are experimental data of the pseudospins $S^+ (S^-)$. {Since the intermediate couplings are non-vanishing, the evolution trajectories will be either inside or outside the Bloch spheres, which mimics the most interesting case of the non-conserving spin current in the BHZ model.}

{Here we show the results of the measuring Berry curvature. Typically, for $g=0$, the Berry curvature is symmetric along $k_x=0$ and $k_y=0$.} At $q=6$, one obtain $M(k)=2B\cos k_x$ with zero frequency shift, and the pseudospins are driven from the north pole to the south pole {(the trajectories are shown in SM \cite{SM})}. By sweeping $k_x$ at each $k_y$ ($q$ from 1 to 11), the Berry curvature in the BZ can be derived from measured $\langle\sigma_{x,y,z}^\tau\rangle$. The theoretical simulation and experimental data of spin Berry curvature $F^s_{k_xk_y}=F^+_{k_xk_y}-F^-_{k_xk_y}$ under the condition of $M=2B$ and $g=0$ are plotted in  Figs. 3(a) and 3(b). \\

{\sl SCNs and topological phase transition.--} The SCN is obtained by integrating the spin Berry curvature over the first BZ with Eq. (\ref{Cs2}).
In Fig. 3(c), we plot the measured $C_s$ and $C_\tau$ versus the parameter $M/2B$ with $g=0$. Theoretically, the BHZ model is in the topological insulating state when  $M/2B>0$ and the trivial insulating state when $M/2B<0$. This topologically nontrivial-to-trivial transition has been confirmed in our experiments, as demonstrated in Fig. 3(c), where  $C_s$ jumps from 1 to 0 at the critical point of $M/2B=0$.  Although relatively large fluctuations are contained in the experimental results of the spin Berry curvature, the measured SCNs $C_s$ (red circles) and CNs $C_\tau$ (black squares and blue diamonds) are stable at 1 and $\pm 1$ respectively,  as shown by Figs. 3(a) and 3(b). This suggests that the SCN is robust against the fluctuations introduced by the controls and measurements.

 We will further demonstrate that using the linear response theory to measure the SCNs is also valid for the model with non-vanishing intermediate coupling $g$. In our experiment, the interaction strength is set to  $g=0.15A$ and the method to prepare the initial state for $g\not= 0$ is addressed in SM \cite{SM}, while the control and measurement procedures are the same as with the case of $g=0$. As shown in Fig. 3(d), in the topological insulating region, the measured $C_\tau$ has large fluctuations, whereas, the SCN  computed from the difference of $C_\tau$ is very stable. This clearly reveals that  $C_\tau$ is not a well-defined topological invariant since it is not even an integer, but the SCN remains valid when intermediate coupling is present between the pseudospins. The above results  meet the rigorous calculations using the U-linked method \cite{Fukui2007}, which confirms the validation of the linear response theory.

As demonstrated in Fig. 3(c) and 3(d), SCNs are robust against variations of the control parameters $M/2B$, and we further show that this robustness remains for the other parameter $g$. $C_s$ and $C_\tau$ versus $g$ for $M=2B$ are plotted in Fig. 3(e).  As the pseudospin gap does not close, SCNs  are stable with $C_s=1$, while  $C_\tau$ numerically calculated with the linear response method fluctuates in this region. Therefore, combining the above results, we observe that SCNs are robust against the parameter variations ($g$ and $M$) as well as random fluctuations (i.e., fluctuations of the measured Berry curvature as shown in Fig. 3(b)).\\

{\sl Conclusions.--}
We have measured topological SCNs  in a simulated BHZ model for the first time.  Our
observations can close the debate whether SCNs can be defined in a
spin non-conserved system \cite{LFu2006,Fukui2007,Prodan2009}. The fully controllable Hamiltonian allows us to investigate other topological models, e.g., the Kane-Mele model \cite{Kane2005Z2}, by employing suitable coupling.  Our work can be extended to other real or artificial atomic
systems, including superconducting qubits, nitrogen-vacancy centers, quantum dots, and
trapped ions \cite{Georgescu2014}.  Since the linear response method used here is experimentally feasible,  our work may lead to the detection of other topological invariants in condensed matter physics and artificial quantum systems. For example, with the spin Hall effect realized with ultracold atoms \cite{Beeler2013,SLZhu2006},  lattice  extensions of our work \cite{Jotzu2014,LBShao2008,Goldman2016,Gianfrat2020}, together with the Feshbach resonance \cite{CChin2010},  may allow  detection of topological invariants for an interacting bosonic or fermionic quantum gas in optical lattices \cite{SLZhu2013,QNiu1985}.
Directly probing these topological invariants is essential for the advance of topological physics and its
quantum simulations.

\bigskip
\begin{acknowledgments}

This work was supported by the Key-Area Research and Development Program of GuangDong Province (Grant No. 2019B030330001),  the Key Project of Science and Technology of Guangzhou (Grant No. 2019050001), the National Key Research and Development Program of China (Grants No. 2016YFA0301800, No. 2016YFA0302800, and No. 2020YFA0309500), and the National Natural Science Foundation of China (Grants No. 12074132, No. 11822403, No. U20A2074, No. 12074180, No. 11804105, No.U1830111, No. 12075090, and No. U1801661).

Q.X.Lv, Y.X.Du, and Z.T.Liang contribute equally to this work.

\end{acknowledgments}


\begin{thebibliography}{99}
\bibitem{Hasan2010} M. Z. Hasan and C. L. Kane, Rev. Mod. Phys. \textbf{82}, 3045
(2010).

\bibitem{XQi2011} X. L. Qi and S. C. Zhang, Rev. Mod. Phys. \textbf{83}, 1057
(2011).
\bibitem{DWZhang2018}D. W. Zhang, Y. Q. Zhu, Y. X. Zhao, H. Yan, and S. L. Zhu, Adv. Phys. \textbf{67}, 253 (2018).

\bibitem{Ozawa2019} T. Ozawa, H. M. Price, A. Amo, N. Goldman, M. Hafezi, L.
Lu, M. Rechtsman, D. Schuster, J. Simon, O. Zilberberg,
and I. Carusotto,
%Topological photonics,
Rev. Mod. Phys.
\textbf{91}, 015006 (2019).


\bibitem{Thouless1982}D. J. Thouless, M. Kohmoto, M. P. Nightingale, and M. den Nijs, Phys. Rev. Lett. \textbf{49}, 405 (1982).

\bibitem{Thouless1983} D. J. Thouless, Phys. Rev. B \textbf{27}, 6083 (1983); Q. Niu and D.
J. Thouless, J. Phys. A \textbf{17}, 2453 (1984).


\bibitem{Haldane1988}F. D. M. Haldane, Phys. Rev. Lett. \textbf{61}, 2015 (1988).

\bibitem{Kane2005} C. L. Kane and E. J. Mele, Phys. Rev. Lett.,
\textbf{95}, 226801 (2005).
%Quantum Spin Hall Effect in Graphene


\bibitem{Kane2005Z2}
%Z2 Topological Order and the Quantum Spin Hall Effect
C. L. Kane and E. J. Mele, Phys. Rev. Lett.,
\textbf{95}, 146802 (2005).

\bibitem{DNSheng2006}
%Quantum Spin-Hall Effect and Topologically Invariant Chern Numbers
D. N. Sheng, Z. Y. Weng, L. Sheng, and F. D. M. Haldane, Phys. Rev. Lett. \textbf{97}, 036808 (2006).

\bibitem{LSheng2005}L. Sheng, D. N. Sheng, C. S. Ting, and F. D. M. Haldane,
Phys. Rev. Lett. \textbf{95}, 136602 (2005).

%\bibitem{DNSheng2006} D. N. Sheng, Z. Y. Weng, L. Sheng, and F. D.
%M. Haldane, Phys. Rev. Lett. \textbf{97}, 036808 (2006).




\bibitem{LFu2006}L. Fu and C. L. Kane, Phys. Rev. B \textbf{74}, 195312 (2006).

\bibitem{Fukui2007}T. Fukui and Y. Hatsugai, Phys. Rev. B \textbf{75}, 121403R (2007).

\bibitem{Prodan2009} E. Prodan, Phys. Rev. B \textbf{80}, 125327 (2009).

\bibitem{SLZhu2013}  S. L. Zhu, Z.-D. Wang, Y.-H. Chan, and L.-M. Duan, Phys. Rev. Lett. \textbf{110}, 075303 (2013).

\bibitem{LSheng2013}L. Sheng, H. C. Li, Y. Y. Yang, D. N. Sheng, and D. Y. Xing, Chin. Phys. B \textbf{22}, 067201 (2013).
%\bibitem{HLi2010}H. C. Li, L. Sheng, D. N. Sheng, and D. Y. Xing, Phys. Rev. B \textbf{82}, 165104 (2010).

\bibitem{Carpentier2015} D. Carpentier, P. Delplace, M. Fruchart, and K. Gawedzki, Phys. Rev. Lett. \textbf{114}, 106806 (2015).

\bibitem{Canonico2019} L. M. Canonico, T. G. Rappoport, and R.B. Muniz, Phys. Rev. Lett. \textbf{122}, 196601 (2019).

\bibitem{CZChen2019} C. Z. Chen, H. Liu, and X. C. Xie, Phys. Rev. Lett. \textbf{122}, 026601 (2019).

\bibitem{LSheng2014} L. Sheng, Progress in Physics, \textbf{34}, 10 (2014).


\bibitem{Bernevig2006} B. A. Bernevig, T. L. Hughes, and S. C. Zhang,
Science, \textbf{314}, 1757 (2006).

\bibitem{Konig2007}M. Konig, S. Wiedmann, C. Brune, A. Roth, H. Buhmann,
L. W. Molenkamp, X. L. Qi, and S. C. Zhang, Science, \textbf{318},
766 (2007).



\bibitem{Hsieh2008}D. Hsieh, D. Qian, L. Wray, Y. Xia, Y. S. Hor, R. J. Cava, and M. Z. Hasan, Nature, \textbf{452}, 979 (2008).

\bibitem{Hsieh2009}D. Hsieh, Y. Xia, L. Wray, D. Qian, A. Pal, J. H. Dil, J. Osterwalder, F. Meier, G. Bihlmayer, C. L. Kane, Y. S. Hor, R. J. Cava, and M. Z. Hasan, Science, \textbf{323}, 919 (2009).

\bibitem{Aidelsburger2015} M. Aidelsburger, M. Lohse, C. Schweizer, M. Atala, J. T. Barreiro, S. Nascimb\`{e}ne, N. R. Cooper,
I. Bloch, and N. Goldman,
%\textit{Measuring the Chern number of Hofstadter bands with ultracold bosonic atoms},
Nat. Phys. \textbf{11}, 162 (2015).

\bibitem{Duca2015} L. Duca, T. Li, M. Reitter, I. Bloch, M. Schleier-Smith, and U. Schneider,
%\textit{An Aharonov-Bohm interferometer for determining Bloch band topology},
Science \textbf{347}, 288 (2015).

 \bibitem{XTan2019}
 %Experimental measurement of quantum metric tensor and related topological phase
%transition with a superconducting qubit
X. Tan, D. W. Zhang, Z. Yang, J. Chu, Y. Q. Zhu, D. Li, X.Yang, S. Song, Z. Han,
Z. Li, Y. Dong, H.F. Yu, H. Yan, S. L. Zhu, and Y. Yu,
Phys. Rev. Lett. \textbf{122}, 210401 (2019).

\bibitem{MYu2020} M. Yu, P. Yang, M. Gong, Q. Cao, Q. Lu, H. Liu, S. Zhang, M. B. Plenio, F. Jelezko, T. Ozawa, N. Goldman, and J. Cai, National Science Review
\textbf{7}, 254  (2020).



\bibitem{XTan2018}
%Topological Maxwell Metal Bands in a Superconducting Qutrit,
    X. Tan, D. W. Zhang, Q. Liu, G. Xue, H. F. Yu,
    Y. Q. Zhu, H. Yan, S. L. Zhu, and Y. Yu,
    Phys. Rev. Lett. \textbf{120}, 130503 (2018).

\bibitem{TLi2016} T. Li, L. Duca, M. Reitter, F. Grusdt, E. Demler, M. Endres, M. Schleier-Smith, I. Bloch and U.
Schneider, Science \textbf{352} 1094 (2016).

\bibitem{Sugawa2018}S. Sugawa, F. Salces-Carcoba, A. R. Perry, Y. C. Yue, and I. B. Spielman, Science \textbf{360}, 1429
(2018).


 \bibitem{Roushan2014} P. Roushan, C. Neill, Y. Chen, M. Kolodrubetz, C. Quintana, N. Leung, M. Fang, R. Barends, B. Campbell,
Z. Chen, B. Chiaro, A. Dunsworth, E. Jeffrey, J. Kelly, A.
Megrant, J. Mutus, P. O'Malley, D. Sank, A. Vainsencher, J.
Wenner, T. White, A. Polkovnikov, A. N. Cleland, and J. M.
Martinis, Nature (London) \textbf{ 515}, 241 (2014).


\bibitem{Schweizer2016}C. Schweizer, M. Lohse, R. Citro, and I. Bloch,
Phys. Rev. Lett. \textbf{117}, 170405 (2016).

\bibitem{Lohse2018} M. Lohse, C. Schweizer, H. M. Price, O. Zilberberg, and I. Bloch,
Nature \textbf{553}, 55 (2018).

\bibitem{SM} See Supplemental Material  for details
on the theoretical model, experimental setup, measurement and data processing.


\bibitem{Gritsev2012} V. Gritseva and A. Polkovnikovb, Proc. Natl. Acad. Sci. \textbf{109}, 17 (2012).

\bibitem{Georgescu2014}I. M. Georgescu, S. Ashhab, and Franco Nori, Rev. Mod. Phys. \textbf{86}, 153 (2014).



%\bibitem{Stuhl2015} B. K. Stuhl, H. I. Lu, L. M. Aycock, D. Genkina, and I. B. Spielman,
%\textit{Visualizing edge states with an atomic Bose gas in the quantum Hall regime},
%Science {\bf349}, 1514 (2015).

%\bibitem{XYang2015} X. C. Yang, D. W. Zhang, P. Xu, Y. Yu, and S. L.
%Zhu, Phys. Rev. A \textbf{91}, 022303 (2015).


\bibitem{Beeler2013} M. C. Beeler, R. A. Williams, K. Jimenez-Garcia, L.
J. LeBlanc, A. R. Perry, and I. B. Spielman, Nature (London)
\textbf{498}, 201 (2013).

\bibitem{SLZhu2006} S. L. Zhu, H. Fu, C. J. Wu, S. C. Zhang, and L.
M. Duan, Phys. Rev. Lett. \textbf{97}, 240401 (2006).


%\bibitem{Lohse2016}
%A Thouless quantum pump with ultracold bosonic atoms in an optical superlattice,
%M. Lohse, C. Schweizer, O. Zilberberg, M. Aidelsburger, and I.
%Bloch, Nat. Phys. {\bf12}, 350 (2016).


%\bibitem{Nakajima2016} %Topological Thouless pumping of ultracold fermions,
%S. Nakajima,  T. Tomita, S. Taie, T. Ichinose, H. Ozawa, L. Wang, M. Troyer, and Y.
% Takahashi, Nat. Phys. {\bf12}, 296 (2016).

\bibitem{Jotzu2014} G. Jotzu, M. Messer, R. Desbuquois, M. Lebrat, T. Uehlinger, D. Greif, and T. Esslinger,
%\textit{Experimental realisation of the topological Haldane model with ultracold fermions},
Nature (London) {\bf515}, 237 (2014).


\bibitem{LBShao2008} L. B. Shao, S. L. Zhu, L. Sheng, D. Y. Xing, and Z.
D. Wang, Phys. Rev. Lett. \textbf{101}, 246810 (2008).

\bibitem{Goldman2016}
%Topological quantum matter with ultracold gases in optical lattices
N. Goldman, J. C. Budich, and P. Zoller, Nature Phys.\textbf{12}, 639 (2016).

\bibitem{Gianfrat2020} A. Gianfrate, O. Bleu, L. Dominici, V. Ardizzone, M. De Giorgi, D. Ballarini, G. Lerario, K. W. West, L. N. Pfeiffer, D. D. Solnyshkov, D. Sanvitto, and G. Malpuech, Nature \textbf{578}, 381 (2020).
%Measurement of the quantum geometric tensor and of the anomalous Hall drift

\bibitem{CChin2010} C. Chin, R. Grimm, P. Julienne, and E. Tiesinga, Rev. Mod. Phys. \textbf{82} 1225 (2010).


\bibitem{QNiu1985} Q. Niu, D.J. Thouless, and Y.S. Wu, Phys. Rev. B \textbf{31}, 3372 (1985).




\end{thebibliography}
\end{document}

\title{Supplemental Materials for\\Measuring the
spin Chern numbers of quantum simulated topological insulators}

\author{Qing-Xian Lv}
\affiliation{Guangdong Provincial Key Laboratory of Quantum Engineering and Quantum Materials, School of Physics and Telecommunication Engineering, South China Normal University, Guangzhou 510006, China}

\affiliation{Guangdong-Hong Kong Joint Laboratory of Quantum Matter, Frontier Research Institute for Physics, South China Normal University, Guangzhou 510006,
China}

\author{Yan-Xiong Du}
\email{yanxiongdu@m.scnu.edu.cn}
\affiliation{Guangdong Provincial Key Laboratory of Quantum Engineering and Quantum Materials, School of Physics and Telecommunication Engineering, South China Normal University, Guangzhou 510006, China}

\author{Zhen-Tao Liang}
\affiliation{Guangdong Provincial Key Laboratory of Quantum Engineering and Quantum Materials, School of Physics and Telecommunication Engineering, South China Normal University, Guangzhou 510006, China}

\author{Hong-Zhi Liu}
\affiliation{Guangdong Provincial Key Laboratory of Quantum Engineering and Quantum Materials, School of Physics and Telecommunication Engineering, South China Normal University, Guangzhou 510006, China}

\author{Jia-Hao Liang}
\affiliation{Guangdong Provincial Key Laboratory of Quantum Engineering and Quantum Materials, School of Physics and Telecommunication Engineering, South China Normal University, Guangzhou 510006, China}

\author{Lin-Qing Chen}
\affiliation{Guangdong Provincial Key Laboratory of Quantum Engineering and Quantum Materials, School of Physics and Telecommunication Engineering, South China Normal University, Guangzhou 510006, China}

\author{Li-Ming Zhou}
\affiliation{Guangdong Provincial Key Laboratory of Quantum Engineering and Quantum Materials, School of Physics and Telecommunication Engineering, South China Normal University, Guangzhou 510006, China}

\author{Shan-Chao Zhang}
\affiliation{Guangdong Provincial Key Laboratory of Quantum Engineering and Quantum Materials, School of Physics and Telecommunication Engineering, South China Normal University, Guangzhou 510006, China}

%\affiliation{ Guangdong Provincial Key Laboratory of Nuclear Science, Institute of Quantum Matter, South China Normal University, Guangzhou 510006, China}
\affiliation{Guangdong-Hong Kong Joint Laboratory of Quantum Matter, Frontier Research Institute for Physics, South China Normal University, Guangzhou 510006,
China}

\author{Dan-Wei Zhang}
\affiliation{Guangdong Provincial Key Laboratory of Quantum Engineering and Quantum Materials, School of Physics and Telecommunication Engineering, South China Normal University, Guangzhou 510006, China}
\affiliation{Guangdong-Hong Kong Joint Laboratory of Quantum Matter, Frontier Research Institute for Physics, South China Normal University, Guangzhou 510006,
China}

\author{Bao-Quan Ai}
\affiliation{Guangdong Provincial Key Laboratory of Quantum Engineering and Quantum Materials, School of Physics and Telecommunication Engineering, South China Normal University, Guangzhou 510006, China}

\affiliation{Guangdong-Hong Kong Joint Laboratory of Quantum Matter, Frontier Research Institute for Physics, South China Normal University, Guangzhou 510006,
China}

\author{Hui Yan}
\email{yanhui@scnu.edu.cn}
\affiliation{Guangdong Provincial Key Laboratory of Quantum Engineering and Quantum Materials, School of Physics and Telecommunication Engineering, South China Normal University, Guangzhou 510006, China}

\affiliation{Guangdong-Hong Kong Joint Laboratory of Quantum Matter, Frontier Research Institute for Physics, South China Normal University, Guangzhou 510006,
China}

\author{Shi-Liang Zhu}
\email{slzhu@scnu.edu.cn}
\affiliation{Guangdong Provincial Key Laboratory of Quantum Engineering and Quantum Materials, School of Physics and Telecommunication Engineering, South China Normal University, Guangzhou 510006, China}

\affiliation{Guangdong-Hong Kong Joint Laboratory of Quantum Matter, Frontier Research Institute for Physics, South China Normal University, Guangzhou 510006,
China}

%\date{\today}
\maketitle

In this Supplemental Material, we provide more detailed discussion
on the properties of the four-level Hamiltonian and experimental details. Especially, the effect of non-adiabatic.

\section{I. Derivation of the four-level Hamiltonian in a cold atomic system}
Here we deduce the effective Hamiltonian of the four-level atomic system shown in Fig. 1(b). Using the bare state basis, the interacting Hamiltonian is given by
\begin{align}\nonumber
&H_b=(\omega_{+E_1}-\omega_{+H_1})|+,E_1\rangle\langle+,E_1|\\\nonumber
&+(\omega_{-E_1}-\omega_{+H_1})|-,E_1\rangle\langle-,E_1|\\\nonumber
&+(\omega_{-H_1}-\omega_{+H_1})|-,H_1\rangle\langle-,H_1|\\\nonumber
&+(\frac{\Omega_1}{2}e^{i\omega_1t}e^{i\varphi_1}|+,E_1\rangle\langle+,H_1|\\\nonumber
&+\frac{\Omega_2}{2}e^{i\omega_2t}e^{i\varphi_2}|-,E_1\rangle\langle-,H_1|\\\nonumber
&+\frac{\Omega_3}{2}e^{i\omega_3t}e^{i\varphi_3}|+,E_1\rangle\langle-,H_1|\\
&+\frac{\Omega_4}{2}e^{i\omega_4t}e^{i\varphi_4}|-,E_1\rangle\langle+,H_1|+H.c.),\label{Hb}
\end{align}
where $\omega_{\pm E_1(H_1)}$ are the energy frequencies of $|\pm,E_1(H_1)\rangle$, and $\Omega_k, \omega_k, \varphi_k$ $(k=1,2,3,4)$ are the Rabi frequencies, frequencies, and phases of the controlling microwaves. { By turning the Hamiltonian to the reference frame $U=e^{-i\omega_1t}|+,E_1\rangle\langle+,E_1|
+|+,H_1\rangle\langle+,H_1|
+e^{-i\omega_4t}|-,E_1\rangle\langle-,E_1|
+e^{-i(\omega_1-\omega_3)t}|-,H_1\rangle\langle-,H_1|$, we can obtain the effective Hamiltonian $H'=i\partial_tU^\dagger U+U^\dagger H_bU$  as}
\begin{equation}
\label{seudospin} H'=\frac{1}{2}\left ( \begin{array}{cccc}
-\Delta_1&\Omega_1e^{-i\varphi_1}&0&\Omega_4e^{-i\varphi_4}\\
\Omega_1e^{i\varphi_1}&\Delta_1&\Omega_3e^{-i\varphi_3}&0\\
0&\Omega_3e^{i\varphi_3}&-\Delta_2&\Omega_2e^{-i\varphi_2}e^{-i\Delta't}\\
\Omega_4e^{i\varphi_4}&0&\Omega_2e^{i\varphi_2}e^{i\Delta't}&\Delta_2
\end{array}\right)
\end{equation}
under the basis $\{|+,E_1\rangle,|+,H_1\rangle,|-,E_1\rangle,|-,H_1\rangle\}$,where
{ $\Delta_1=\omega_1-(\omega_{+E_1}-\omega_{+H_1}), \Delta_2=\omega_2-(\omega_{-E_1}-\omega_{-H_1}), \Delta_3=\omega_3-(\omega_{+E_1}-\omega_{-H_1}), \Delta_4=\omega_4-(\omega_{-E_1}-\omega_{+H_1})$, $\Delta'=\omega_1+\omega_2-\omega_3-\omega_4=\Delta_1+\Delta_2-\Delta_3-\Delta_4$}. It can be found that the Hamiltonian is time-independent  by choosing $\Delta'=0$, which induces $\Delta_3+\Delta_4=\Delta_1+\Delta_2$. Therefore, if one sweeps $\omega_1$ and $\omega_2$ simultaneously ($\Delta_1=\Delta_2=\Delta$), the frequencies $\omega_{3,4}$ should also be swept simultaneously, i.e. keeping $\Delta_3=\Delta_4=\Delta$. In our case, by choosing $\Omega_3=\Omega_4=g,$ $\Delta_1=\Delta_2=B_z,$ $\Omega_1=\Omega_2=\Omega=\sqrt{B_x^2+B_y^2},$ $\varphi_1=\varphi_2=\varphi=\arctan{B_y/B_x}$, $\varphi_3=\varphi_4=0$, and $\Delta'=0$, the Hamiltonian (1) in the main text can be derived. As can be seen, controlling the relative phases between the Rabi frequencies plays a key role in the experiments.\\

\setcounter{figure}{0}
\renewcommand{\thefigure}{S1}
\begin{figure}[ptb]
\begin{center}
\includegraphics[width=8cm]{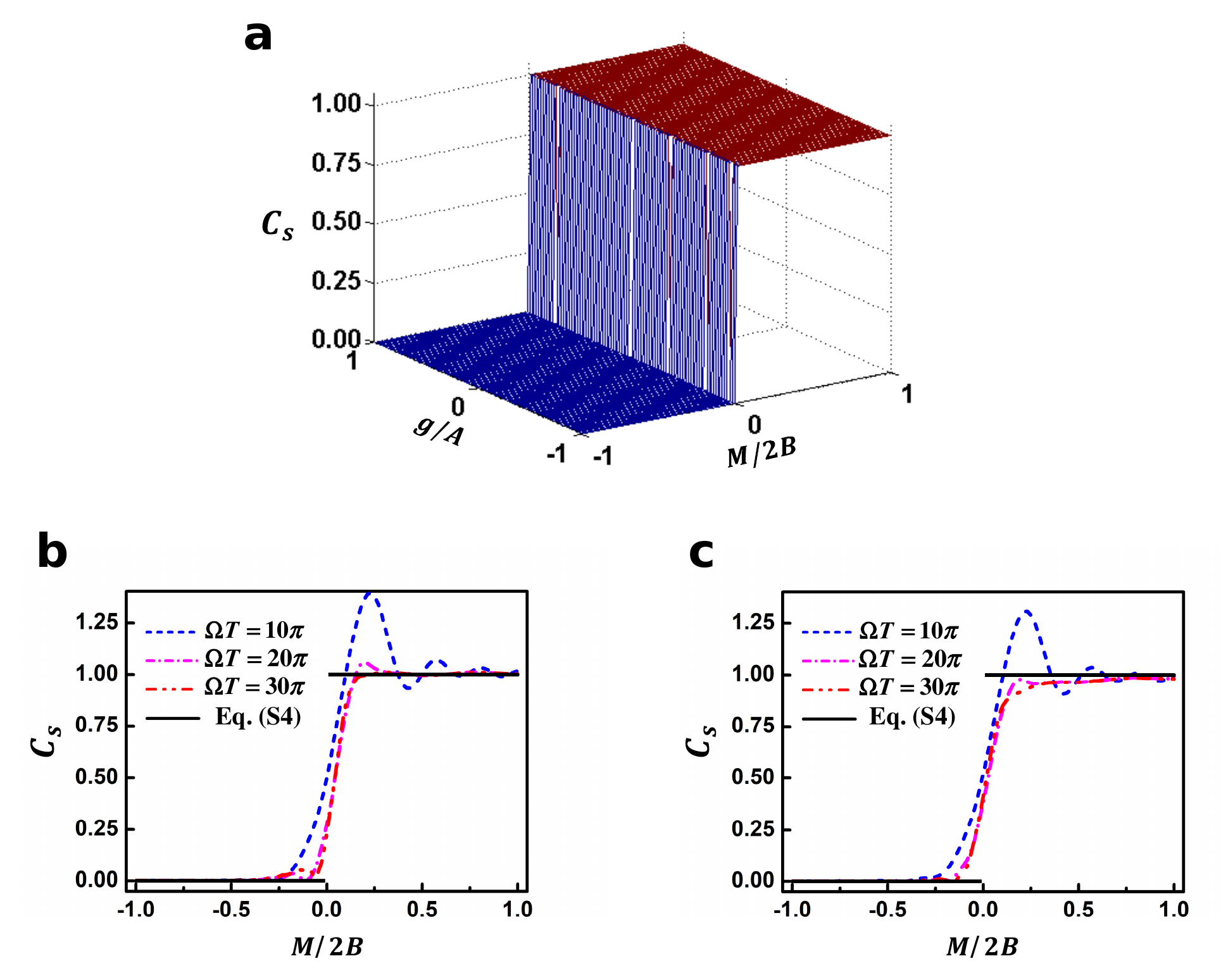}
\caption{(a) SCNs calculated using the U-link method (Eq. (\ref{Cs1}))  versus parameters $g$ and $M$. { (b,c) SCNs calculated using the linear response theory (Eq. (5) and (4))  and U-link method (Eq. (\ref{Cs1})) with $g=0$ in (b) and $g=0.15A$ in (c) }. The SCNs calculated using the  linear response theory jump more and more steeply at $M/2B=0$ as $\Omega T$ increases.
}
\end{center}
\end{figure}

%%%%%%%%%%%%%%%%%%%%%%%%%%%%%%%%%%%%%%%%%%%%%%%%%%%%%%%%

\section{II. Calculation of SCNs using the U-link method}
The SCNs that characterize the topology of the BHZ model can be calculated by generalizing  the U-link method proposed in Ref. \cite{Fukui2007}. Firstly, we discretize the BZ as $(k^x_r=-\pi+2\pi
r/{R},k^y_{n}=-\pi+2\pi n/{N})$, $r=0,1,2,\cdots,R; n=0,1,2,\cdots,N $, and
then numerically diagonalize the discretized Hamiltonian
$H(k^x_r,k^y_{n})$ in Eq. (1) in the main text. We obtain the
normalized eigenfunctions $|\phi_j\rangle$ ($j=1,2,3,4$) related
to the eigenenergies $\mathcal{E}_j$ with the order $\mathcal{E}_4 \geq \mathcal{E}_3 \geq
\mathcal{E}_2\geq \mathcal{E}_1$. We  then decomposite the two eigenfunctions with
lowest eigenenergies in two sectors of the pseudospin operator
$I \otimes \sigma_z$. We thus derive a $2\times 2$ matrix given by
%%%%%%%%%%%%%%%%%%%%%%%%%%%%%%%%%%%%%%%%%%%%%%%%%%%%%%%%%%%%%%%%%%%%%%%%%%%%%%%%%%%%%%%%%%%%%%%%%%%%%%%
\begin{equation}
\label{Seudospin} H^s=\left ( \begin{array}{cc} \langle \phi_1
|I \otimes \sigma_z|\phi_1\rangle & \langle \phi_1 |I \otimes
\sigma_z|\phi_2\rangle \\  \langle \phi_2 |I \otimes
\sigma_z|\phi_1\rangle & \langle \phi_2 |I \otimes
\sigma_z|\phi_2\rangle
\end{array}\right).\end{equation}
%%%%%%%%%%%%%%%%%%%%%%%%%%%%%%%%%%%%%%%%%%%%%%%%%%%%%%%%%%%%%%%%%%%%%%%%%%%%%%%%%%%%%%%%%%%%%%%%%%%%%%%%%%%%
We further diagonalize the reduced Hamiltonian $H^s$ to get two eigenvalues $\mathcal{E}_\pm$, which
can be used to define the pseudospin gap
%%%%%%%%%%%%%%%%%%%%%%%%%%%%%%%%%%%%
%%%%%%%%%%%%%%%%%%%%%%%%%%%%%%%%%%
$\Delta_s
\equiv \text{Min}|\mathcal{E}_{+}-\mathcal{E}_{-}|
$
%%%%%%%%%%%%%%%%%%%
%%%%%%%%%%%%%%%%%%%%
and its related normalized eigenfunctions
$|\psi_{\tau}\rangle=(a^{\tau},b^{\tau})^\text{Tr}$. The SCN is
well defined if the pseudospin gap $\Delta_s>0$.  Let
us denote $|\Psi^\tau (k^x_r,k^y_{n})\rangle=a^\tau
|\phi_1(k^x_r,k^y_{n})\rangle+ b^\tau |\phi_2(k^x_r,k^y_{n})\rangle$, where $(a^\tau, b^\tau)$ are the two eigenfunctions of the reduced Hamiltonian $H^s$. We can then define a
U-link variable from the wavefunction as follows
%%%%%%%%%%%%%%%%%%%%%%%%%%%%%%%%%%%%%%%%
%%%%%%%%%%%%%%%%%%%%%%%%%%%%%%%%%%%%%%%%
\begin{eqnarray*}
&& U_{r,n}^\tau = \langle\Psi^\tau
(k^x_{r},k^y_{n})|\Psi^\tau
 (k^x_{r+1},k^y_{n})\rangle,\\
&&U_{r+1,{n}}^\tau = \langle\Psi^\tau
(k^x_{r+1},k^y_{n})|\Psi^\tau (k^x_{r+1},k^y_{n+1})\rangle,\\
&&U_{r+1,n+1}^\tau =\langle\Psi^\tau
(k^x_{r+1},k^y_{n+1})|\Psi^\tau (k^x_{r},k^y_{n+1})\rangle,\\
&&U_{r,n+1}^\tau =\langle\Psi^\tau
(k^x_{r},k^y_{n+1})|\Psi^\tau
(k^x_{r},k^y_{n})\rangle.
\end{eqnarray*}
%%%%%%%%%%%%%%%%%%%%%%%%%%%%%%%%%%%%%
%%%%%%%%%%%%%%%%%%%%%%%%%%%%%%%%%%%%%
The SCN can then be efficiently obtained by the following form:
\begin{equation}
\label{Cs1} C_s=\sum_{r n\tau} \left[-\frac{\tau}{2} \ln\left(
U_{r,n}^\tau U_{r+1,n}^\tau U_{r+1, n+1}^\tau
U_{r,n+1}^\tau\right)\right]
\end{equation}
under the condition that there is a finite pseudospin gap
$\Delta_s>0$.
%The  expression of $C_s$ in Eq. (\ref{Cs1}) is
%failed when the pseudospin is gapless, i.e., $\Delta_s=0$.
The SCNs calculated using this method are plotted in Fig. S1(a).\\

%%%%%%%%%%%%%%%%%%%%%%%%%%%%
%\setcounter{figure}{0}
%\renewcommand{\thefigure}{6}
\setcounter{figure}{0}
\renewcommand{\thefigure}{S2}
\begin{figure*}[ptb]
\begin{center}
\includegraphics[width=12cm]{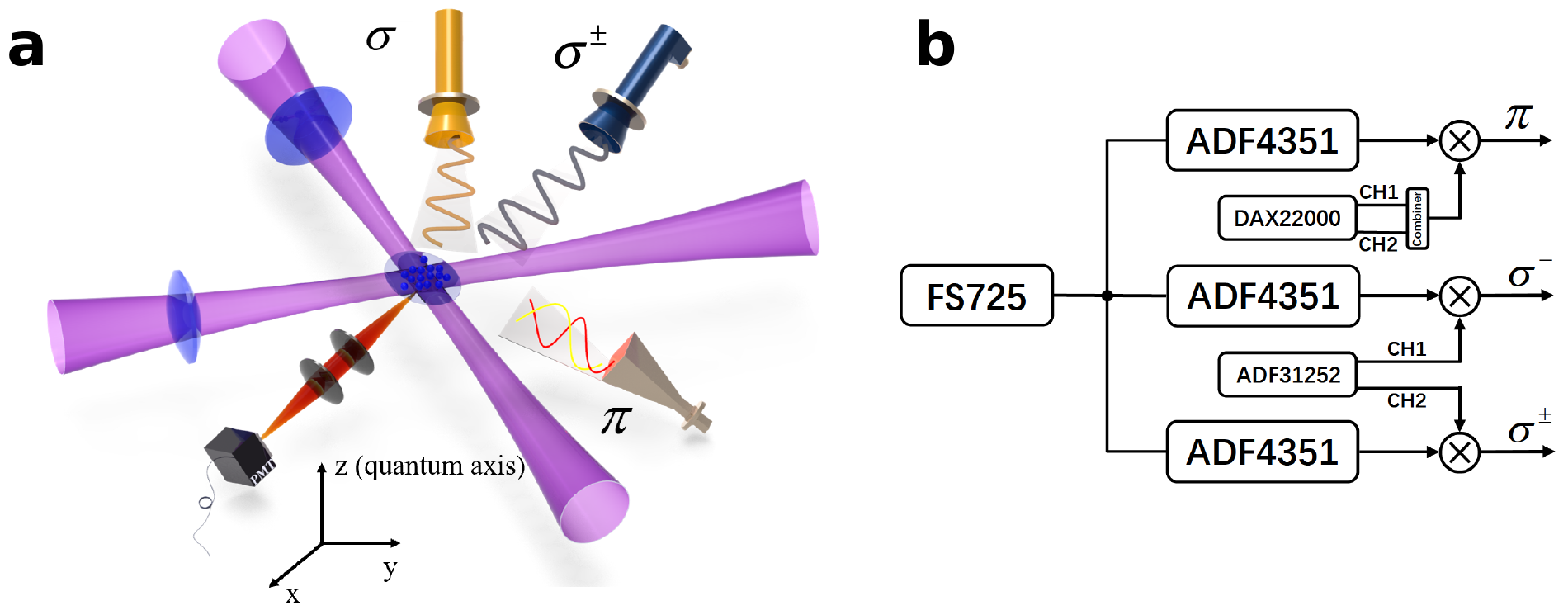}
\caption{Experimental setup. (a) Cold atoms trapped in a crossed optical dipole trap. A quantization axis along the z axis is used to split the magnetic sublevels for addressable manipulation. Two microwaves emitted from a linear-polarization horn antenna with $\pi$-transition are used to control pseudospins, while another microwave emitted from a circular-transition antenna (labelled $\sigma^\pm$ in the figure)  is used to produce the intermediate coupling between the blocks $S_{+}$ and $S_{-}$. An additional microwave with $\sigma^-$-transition is applied to detect the four-level population.  (b) Electronic circuit of the microwave system. To keep phase coherence between the different microwaves, all eigen sources are synchronized to the same atomic clock. FS725: atomic clock. ADF4351: eigen microwave sources. DAX22000: two channels arbitrary waveform generator (CH1 and CH2). ADF31252: arbitrary function generator. By mixing the signals from eigen sources and radio sources, the arbitrary time-varying amplitude, frequency, and phase microwave field can be achieved.}
\end{center}
\end{figure*}
%%%%%%%%%%%%%%%%%%%%%%%%%%%%%%%%%%%%%%%%%%%%%%%%%%%%%%%%%%%%%%%%%%%%%%%%%%%%%%%%%%%%%%%%%%
%\setcounter{figure}{0}
%\renewcommand{\thefigure}{7}
\setcounter{figure}{0}
\renewcommand{\thefigure}{S3}
\begin{figure*}[ptb]
\begin{center}
\includegraphics[width=14cm]{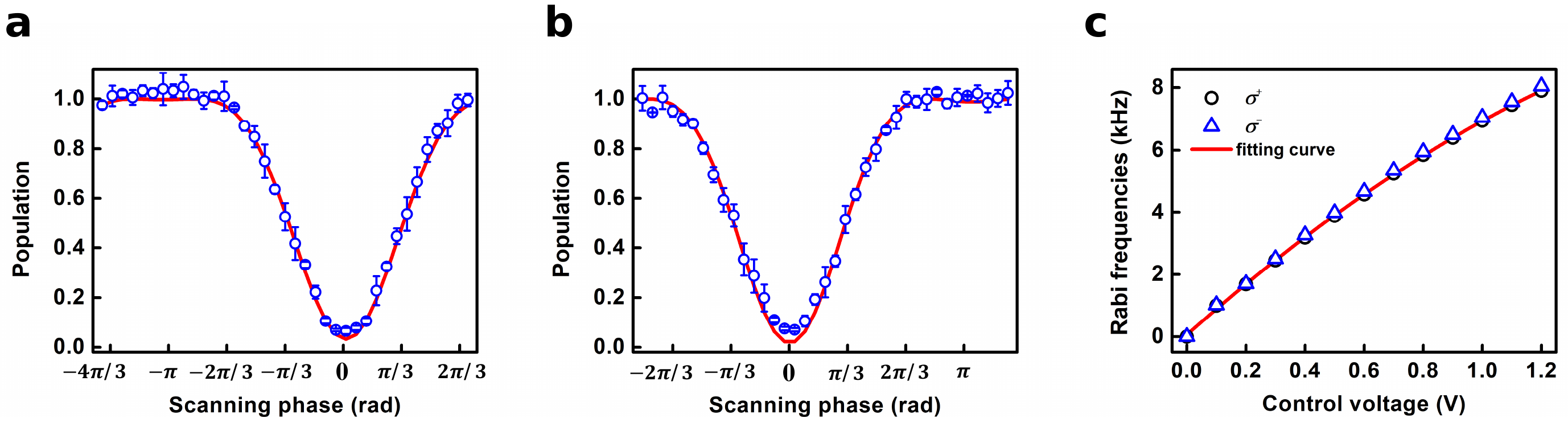}
\caption{Calibrating the phases and amplitudes of the microwaves. (a)(b) Population difference by scanning the relative phases between two  microwave pulses. (a) $\sigma^\pm$ versus $\pi$. (b) $\sigma^-$ versus $\pi$. Red solid lines: fitting curves with the form $f(\phi)=\sqrt{\Omega_a^2+\Omega_b^2+2\Omega_a\Omega_b\cos{\phi}}$, where $\Omega_a$ and $\Omega_b$ are the Rabi frequencies and $\phi$ is the relative phase. Blue dots with error bars: experimental data. (c) Rabi frequencies of the intermediate coupling $g$ controlled with $\sigma^\pm$ . Black circles: $\sigma^+$. Blue triangles: $\sigma^-$. Red solid line: fitting curve with a polynomial. The Rabi frequencies of $\sigma^+$ and $\sigma^-$ can be controlled to be equal at different control voltages.}
\end{center}
\end{figure*}

%%%%%%%%%%%%%%%%%%%%%%%%%%%%%%%%%%%%%%%%%%%%%%%%%%%%%%%%

\section{III. Measurement of SCNs using the linear response theory}
Here we describe the method to measure the SCN in our system. { Following the proposal outlined in Ref. \cite{Gritsev2012}, the generalized force for the total system can be derived by $\langle f_{k_y}\rangle=(\langle \Psi (t)| \partial_{k_y} H_{BHZ}
|\Psi(t)\rangle-\langle \Psi (0)| \partial_{k_y} H_{BHZ}
|\Psi(0)\rangle)/v_{k_x},$ provided that the system
parameters are changed slowly enough that the second order terms $O
(v_{k_x}^2)$ are negligible. Here $v_{k_x}$ is the
rate of change for the parameter $k_x$. In the general cases, the coupling $g$ is non-vanishing, nevertheless, is time-independent. Under this condition and  using the parameterizations in Eq.(2) in the main text, we can obtain $$\langle \partial_{k_y} H_{BHZ} \rangle=A\cos k_y\langle \Gamma_y\rangle+2B\cos k_y\langle \Gamma_z\rangle,$$ where
\begin{equation}
\label{gammamatrixs} \Gamma_x=\left ( \begin{array}{cccc}
0&1&0&0\\
1&0&0&0\\
0&0&0&1\\
0&0&1&0
\end{array}\right),
\Gamma_y=\left ( \begin{array}{cccc}
0&i&0&0\\
-i&0&0&0\\
0&0&0&i\\
0&0&-i&0
\end{array}\right),
\end{equation}
\nonumber \begin{equation}
\label{gammamatrixs1} \Gamma_z=\left ( \begin{array}{cccc}
1&0&0&0\\
0&-1&0&0\\
0&0&1&0\\
0&0&0&-1
\end{array}\right).
\end{equation}
As non-diagonal blocks in Eq.(S5) for different pseudospins are vanishing, the total generalized force can be achieved by separately measuring the response of each pseudospin by nonadiabatically
manipulating the Hamiltonian $H_{\tau} (k_x, k_y)$,
which leads to a generalized force $\langle f_{k_y}^\tau \rangle=\langle
\eta^\tau(0)|f_{k_y}^\tau|\eta^\tau(0)\rangle+v_{k_x}
F_{k_xk_y}^\tau$, where $|\eta^\tau(0)\rangle$ is the initial state of the pseudospin $\tau$.
%Therefore, the Berry curvature
%$F_{k_xk_y}^\tau$ may be derived from the linear response
%\cite{Gritsev2012}.

We prepare the initial state given by $|\Psi(0)\rangle=\left ( \begin{array}{c}
|\eta^+(0)\rangle\\
|\eta^-(0)\rangle
\end{array}\right)$, and $\left ( \begin{array}{c}
1\\
0
\end{array}\right)\otimes|\eta^+(0)\rangle=|\phi_1(0)\rangle, \left ( \begin{array}{c}
0\\
1
\end{array}\right)\otimes|\eta^-(0)\rangle=|\phi_2(0)\rangle$, where $|\eta^\tau(t)\rangle$ is the wave function defined in pseudospin $S_\tau$  and $|\phi_{1,2}(0)\rangle$ are the two lower normalized eigenstates of the Hamiltonian $H_{BHZ}(k_x,k_y)$ at $k_x=-\pi$ and $k_y=k_y(q)$ for a specific $q$. As also mentioned in the main text, $\langle \Psi (t) |\Psi(t)\rangle=2$ since it is more convenient if the wavefunction of each pseudospin itself is normalized to one.} The reason for the equal superposition of two
lower eigenstates $|\phi_{1,2}(0)\rangle$ is that this case is corresponding to the two lower bands of the BHZ model are all occupied. Then we sweep $k_x$  with $k_x(t)=v_{k_x}t-\pi$ with the time $t\in [0,T]$, stopping the ramp at
various times $t_{meas} \leq T$ to perform tomography. At each
$t_{meas}$, we measure the generalized force
%%%%%%%%%%%%%%%%%%%%%%%%%%%%%
%%%%%%%%%%%%%%%%%%%%%%%%%%%%%%%%%
$\langle   f_{k_y}^\tau\rangle= -\langle \eta^\tau (t)| \partial_{k_y} H_\tau
|\eta^\tau(t)\rangle$. Here $|\eta^\tau(t)\rangle$ is the projected wavefunction that the total wavefunction $|\Psi(t)\rangle$ projects onto the pseudospin $\tau$ and $|\Psi(t)\rangle=\hat{P}\exp[-i/\hbar\int H_{BHZ} (t) dt]$.
 Therefore, the Berry curvature is given as the leading order correction to adiabatic
manipulation by  $F_{k_xk_y}^\tau=(\langle f_{k_y}^\tau\rangle-\langle f_0^\tau\rangle)/v_{k_x}$ with $\langle f_0^\tau\rangle=\langle
\eta_0^\tau|f_{k_y}^\tau|\eta_0^\tau\rangle,$ which is Eq. (5) in the main text.
%%%%%%%%%%%%%%%%%%%%%%%%%%%%%%%%%%%%%%%%%%%%%%%%%%%%%%%%%%%%%

The initial state $|\Psi(0)\rangle=(|\phi_1\rangle+|\phi_2\rangle)$  can be further parameterized as $|\Psi(0)\rangle=(\alpha|\eta^+(0)\rangle e^{i\varphi_{+-}},\beta|\eta^-(0)\rangle)^{\mathrm{Tr}}$, where $|\eta^+(0)\rangle=(-\sin(\theta_+/2)e^{i\varphi_+},\cos(\theta_+/2))$ and $|\eta^-(0)\rangle=(-\sin(\theta_-/2)e^{i\varphi_-},\cos(\theta_-/2))$. When $g=0$, the initial state is $|\Psi(0)\rangle=(0,1,0,1)^{\mathrm{Tr}}$ in the basis $\{|+,E_1\rangle,|+,H_1\rangle,|-,E_1\rangle,|-,H_1\rangle\}$.  When $g \neq 0$, the initial state parameterized by the set $\{\alpha,\beta,\theta_{\pm},\varphi_\pm\}$ should be determined numerically by solving the Schrodinger equation $H_{BHZ}(t=0)|\phi_k\rangle=E|\phi_k\rangle$. Experimentally, the atoms are first pumped to $|-, E_1\rangle$ by optical pumping. We then design the pulses according to the known parameters $\{\alpha,\beta,\theta_{\pm},\varphi_\pm\}$ . The population is transferred to $|-, H_1\rangle$ by a $\pi$ pulse  blowing the left atoms in $|-, E_1\rangle$. After that, a $\sigma^-$-transition microwave pulse determined by the parameters $\{\varphi_{+-}, \alpha, \beta\}$ transfers the population from $|-, H_1\rangle$ to $|+, E_1\rangle$. Furthermore, two $\pi$-transition pulses determined by the parameters $\{\theta_\pm.\varphi_\pm\}$ are applied to $S_\pm$. After these manipulations, we can create the desired initial state.\\

In Fig. S1(b)(c), the SCNs of the BHZ model obtained by the linear response theory {with $g=0$ and $g=0.15A$, respectively} are plotted. A smaller $\Omega T$ will drag the evolution trajectory further from the adiabatic case, while the topological transition will jump gently. A larger $\Omega T$ will induce a steeper topological transition but suffer more from the decoherence. As can be seen from Fig. S1(b), the SCNs  calculated using the linear response theory agree well with the rigorous ones derived from the U-link method. In our experiments, we choose $\Omega T=24\pi$.\\

%%%%%%%%%%%%%%%%%%%%%%%%%%%%%%%%%%%%%%%%%%%%%%%%
\setcounter{figure}{0}
\renewcommand{\thefigure}{S4}
\begin{figure*}[ptb]
\begin{center}
\includegraphics[width=16cm]{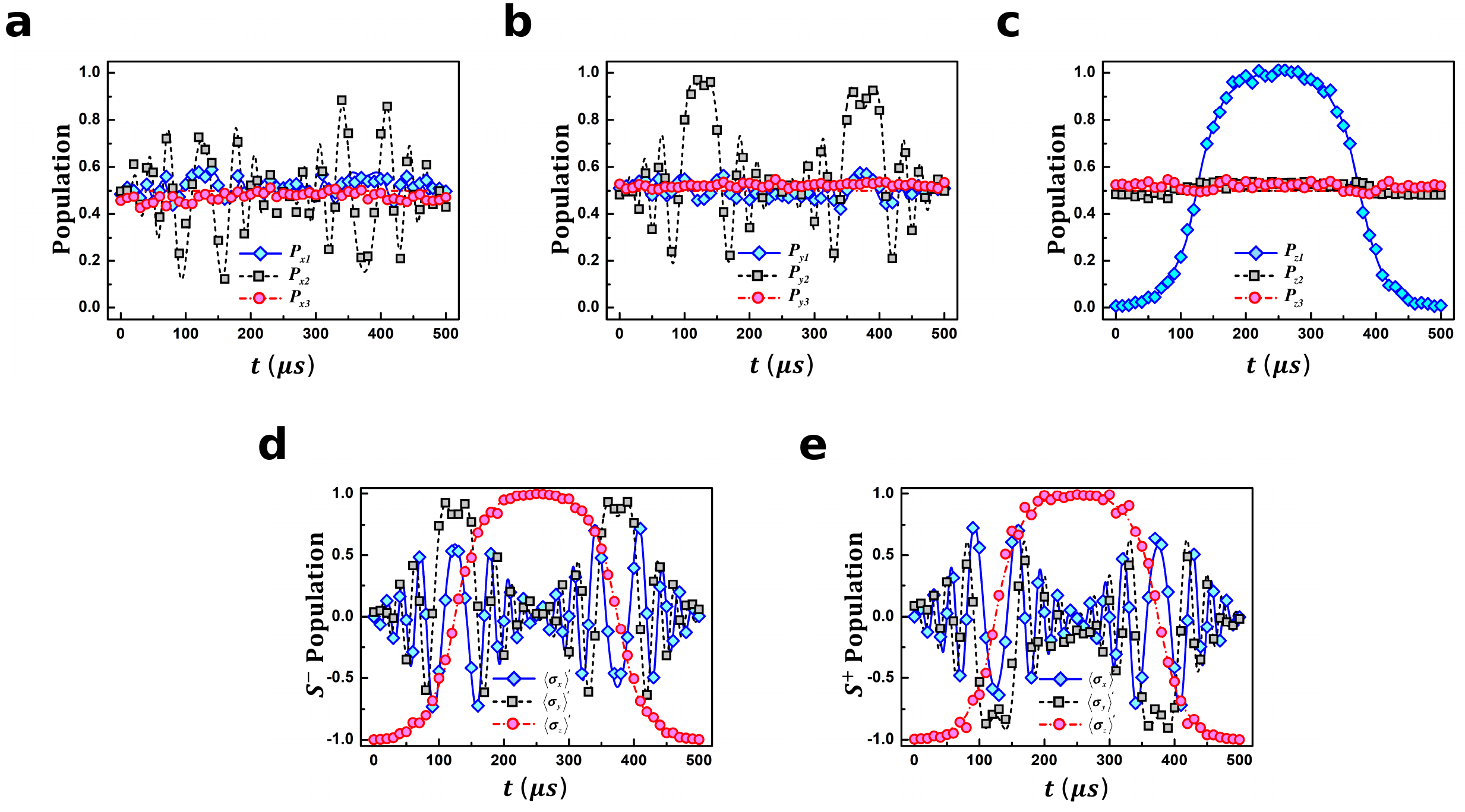}
\caption{Population dynamics of the four-level system in a linear response. (a)(b)(c) The four level population $P_{x_l}$, $P_{y_l}$, $P_{z_l}, l=1,2,3,$ in the measurement procedure. Lines: theoretical curves calculated by solving the Sch\"{o}rdinger equation with the Hamiltonian (\ref{seudospin}). Diamonds, squares, and circles: experimental data. (d)(e) Population dynamics of pseudospins $S_-$ and $S_+$ with the data in (a), (b), and (c). These data are connected to those in Fig. 2(c) by rotation (\ref{Rotation}). Lines: theoretical curves calculated from the Sch\"{o}rdinger equation with Hamiltonian (\ref{seudospin}). Diamonds, squares, and circles: experimental data. Parameters: $M=2B$ and $g=0$.}
\end{center}
\end{figure*}

%%%%%%%%%%%%%%%%%%%%%%%%%%%%%%%%%%%%%%%%%%%%%%%%%%%%%%%%

%%%%%%%%%%%%%%%%%%%%%%%%%%%%%%%%%%%%%%%%%%%%%%%%
\setcounter{figure}{0}
\renewcommand{\thefigure}{S5}
\begin{figure*}[ptb]
\begin{center}
\includegraphics[width=16cm]{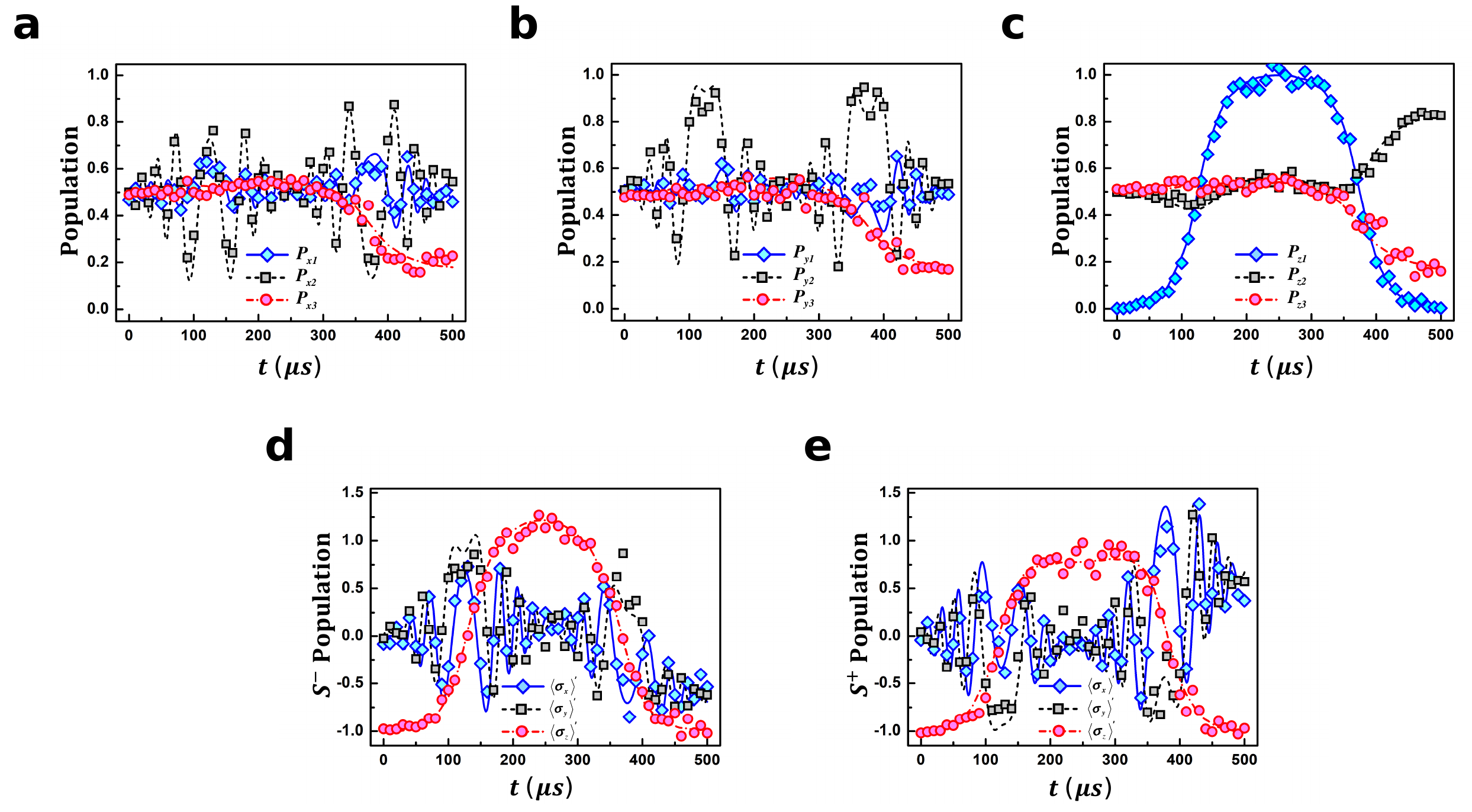}
\caption{Population dynamics of the four-level system in a linear response. (a)(b)(c) The four level population $P_{x_l}$, $P_{y_l}$, $P_{z_l}, l=1,2,3,$ in the measurement procedure. Lines: theoretical curves calculated by solving the Sch\"{o}rdinger equation with the Hamiltonian (\ref{seudospin}). Diamonds, squares, and circles: experimental data. (d)(e) Population dynamics of pseudospins $S_-$ and $S_+$ with the data in (a), (b), and (c). These data are connected to those in Fig. 2(c) by rotation (\ref{Rotation}). Lines: theoretical curves calculated from the Sch\"{o}rdinger equation with Hamiltonian (\ref{seudospin}). Diamonds, squares, and circles: experimental data. Parameters: $M=2B$ and $g=0.15A$.}
\end{center}
\end{figure*}

%%%%%%%%%%%%%%%%%%%%%%%%%%%%%%%%%%%%%%%%%%%%%%%%%%%%%%%%

%%%%%%%%%%%%%%%%%%%%%%%%%%%%%%%%%%%%%%%%%%%%%%%%
\setcounter{figure}{0}
\renewcommand{\thefigure}{S6}
\begin{figure}[ptb]
\begin{center}
\includegraphics[width=8cm]{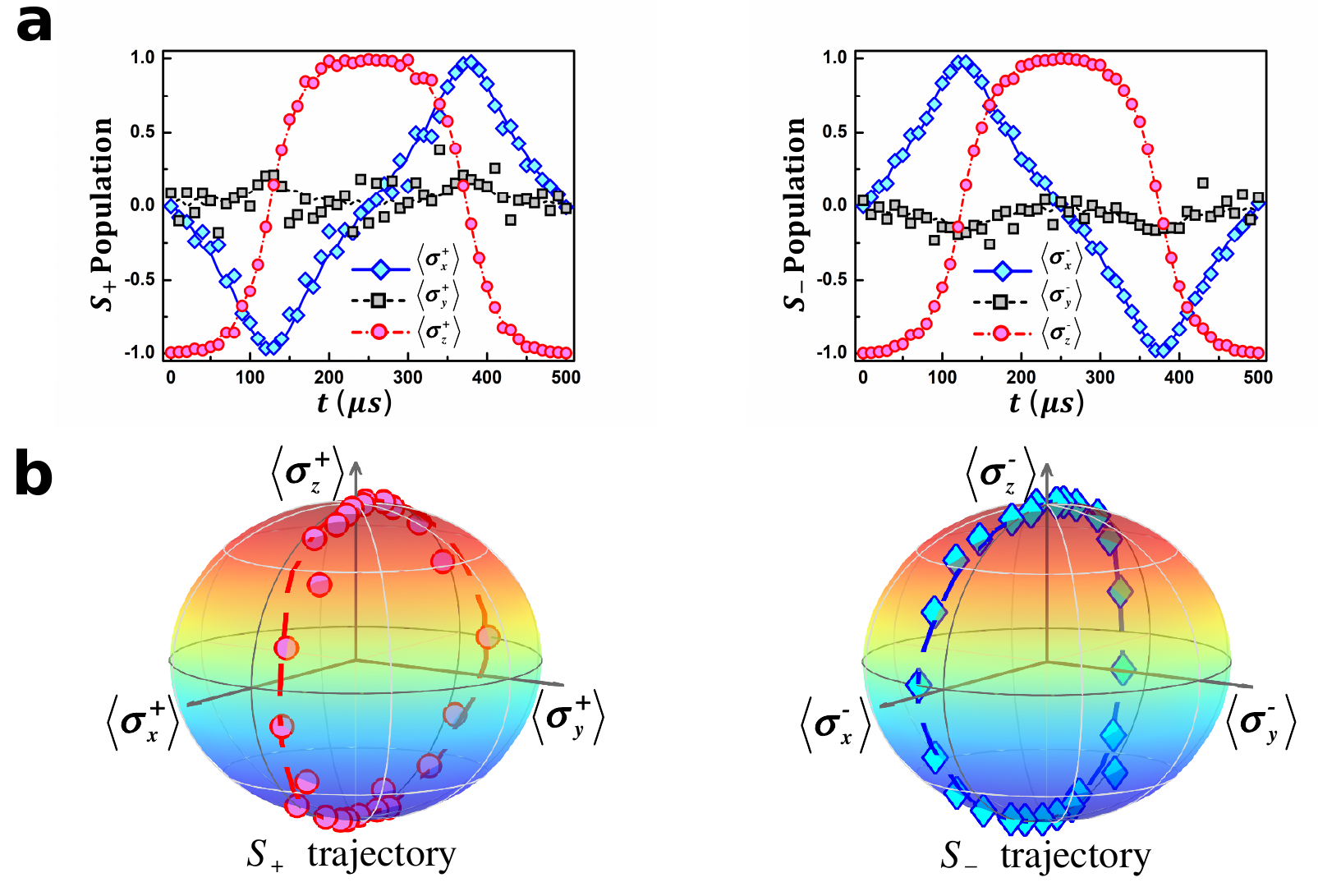}
\caption{{(a) Tomography of pseudospin $S_\pm$ which corresponds to  parameter quenches along the line denoted by points $P$ and $Q$ in Fig.2(a). Dashed and solid lines are  theoretical simulations with the Sch\"{o}dinger equation using  Hamiltonian (1). The evolving four-level state is divided into two pseudospins $S_\pm$. Diamonds, squares, and circles are  experimental data. (b) Trajectories of $S_\pm$ in (a). Dashed lines are  theoretical simulations, and diamonds and circles are experimental data. Parameters: $M=2B $ and $g=0$}}
\end{center}
\end{figure}

%%%%%%%%%%%%%%%%%%%%%%%%%%%%%%%%%%%%%%%%%%%%%%%%%%%%%%%%

\section{IV. Experimental setup}
The schematic diagram of  the experimental setup is shown in Fig. S2(a). Cold atoms are trapped in an optical dipole trap. Three microwave horn antennas are used to couple different levels of $^{87}$Rb. The linear polarization horn antenna (labelled $\pi$)  with two different frequencies emits linear polarization microwaves to control $S^+$ and $S^-$. The circular polarization horn antenna (labelled $\sigma^\pm$) emits circular polarization microwaves with $\sigma^+$-transition. Due to the non-purity of the antenna and the back reflection from the environment, $\sigma^-$-transition is also produced from the circular polarization microwaves. Therefore, the $\sigma^\pm$-transition of the intermediate coupling can be realized by only one circular polarization horn antenna. Another circular polarization horn antenna (labelled $\sigma^-$) is used for state preparation and detection, and population is measured with normalization by photomultiplier tubes (PMT). In Fig. S2(b), we show a schematic diagram of the electrical connection of the microwave device. The microwave eigen sources are all generated by frequency synthesizers (Analog devices, ADF4351) that are connected to the same atomic clock. The $\pi$ horn antenna emits microwaves generated by mixing the radio frequencies (generated by Wavepond, DAX22000-8M) and  eigen frequencies. The $\sigma^-$ and $\sigma^\pm$ horn antennas emit microwaves generated by mixing the radio frequencies (generated by Tektronix, ADF31252) and  eigen frequencies.\\

%%%%%%%%%%%%%%%%%%%%%%%%%%%%%%%%%%%%%%%%%%%%%%%%%%%%%%%%

\section{V. Calibration of the relative phases}
To realize the BHZ model, relative phases among the microwaves should be carefully calibrated. In the experiments, the phase differences may be caused by transport pathes, electronic delays, and the radio frequencies used to initialize the phases of the microwaves. We detect the relative phases of the microwaves by detecting the interference between the Rabi frequencies of the corresponding microwaves. We prepare the system to state $|1,0\rangle$, and adjust two of the microwaves to be resonant with $|1,0\rangle$ and $|2,0\rangle$  simultaneously. Given an operation time, the induced population will be influenced by the synthetic Rabi frequency that is determined by the phase difference between the Rabi frequencies of the microwaves. The relative phases between the microwaves can thus be extracted from the population measurements. In Fig. S3(a) and (b), the relative phases between $\pi$-transition microwave and one of the $\sigma^\pm (\sigma^-)$-transitions are plotted, where the lowest population indicates zero phase difference. The blue circles with error bars are the experimental data and the red solid curves are theoretical curves with the form $f(\phi)=\sqrt{\Omega_a^2+\Omega_b^2+2\Omega_a\Omega_b\cos{\phi}}$, where $\Omega_a$ and $\Omega_b$ are the Rabi frequencies, and $\phi$ is the relative phase. Finally, we fix the phase of one microwave (i.e., $\pi$-transition microwave) and adjust the phases of $\sigma^\pm$, $\sigma^-$-transition microwaves to fulfill the phases conditions. Note that the two channels of $\pi$-transition microwaves generate from the same resource and transport along the same pathes, the phases of the two microwaves are  automatically the same.\\

%%%%%%%%%%%%%%%%%%%%%%%%%%%%%%%%%%%%%%%%%%%%%%%%%%%%%%%%

\section{VI. Calibration of Rabi frequencies of the intermediate couplings}
We apply intermediate couplings between the pseudospins to simulate the physics of the non-conversing spin population in the BHZ model. $\sigma^+$ and $\sigma^-$-transition in the intermediate coupling should be equal according to the form of Hamiltonian (1), which can be realized by regulating the position of the $\sigma^\pm$ horn antenna. We prepare the system to state $|1,-1\rangle$ and $|1,0\rangle$ and measure the Rabi oscillation caused by  $\sigma^+$ and $\sigma^-$, respectively.  According to the measured Rabi frequencies, we carefully modify the position of the $\sigma^\pm$ horn antenna until the $\sigma^+$ and $\sigma^-$-transitions in the intermediate coupling are equal. The $\sigma^+$ and $\sigma^-$-transition data are plotted in Fig. S3(c), which shows that  the Rabi frequencies of $\sigma^+$ (black circles) are equal to that of $\sigma^-$ (blue triangles) as the control voltage changes, confirming that good intermediate coupling can be realized.\\

%%%%%%%%%%%%%%%%%%%%%%%%%%%%%%%%%%%%%%%%%%%%%%%%%%%%%%%%

\section{VII. Measurement and data processing}
To obtain the spin Berry curvature and  thus the SCNs, the expectation values  $\langle\sigma_{n}^\tau\rangle$ ($n=x,y,z$) of the pseudospins should be measured. The values  $\langle\sigma_{n}^\tau\rangle$ in Eq. (5) are defined by the reference frame rotating with  microwave frequencies;  the measured values in our experiments are $\langle\sigma_{n}^\tau\rangle^\prime$, which are related to the lab frame rotating with atomic level frequencies. The expectation values from the two frames differ with a phase $\varphi_0=\int_0^TB_zdt$ along the $z$ axis, i.e., \cite{Mgong2016}
\begin{equation}
\label{Rotation} \left ( \begin{array}{c}
\langle\sigma_x^\tau\rangle\\
\langle\sigma_y^\tau\rangle\\
\langle\sigma_z^\tau\rangle
\end{array}\right)   =\left ( \begin{array}{ccc}
\cos\varphi_0 & -\sin\varphi_0 & 0\\
\sin\varphi_0 & \cos\varphi_0 & 0\\
0 & 0 & 1
\end{array}\right)\left ( \begin{array}{c}
\langle\sigma_x^\tau\rangle'\\
\langle\sigma_y^\tau\rangle'\\
\langle\sigma_z^\tau\rangle'
\end{array}\right) .
\end{equation}

By pumping populations in the upper levels ($|\pm E_1\rangle$) in $S_\pm$ to the excited states, i.e., $|F'=3\rangle$ of $5^2\mathrm{P}_{3/2}$ of $^{87}$Rb, we can detect this population by collecting the fluorescence. We label the population in level $|\pm E_1\rangle$ and $|\pm H_1\rangle$ as $P_{\pm,E_1}$ and $P_{\pm,H_1}$, respectively.  To detect $\langle\sigma_z^\tau\rangle'$, we adopt the following three steps. (i) Directly project measurement of the upper levels ($|\pm E_1\rangle$) in $S_\pm$. The population stay in $|\pm E_1\rangle$ will be $P_{+,E_1}, P_{-,E_1}$ and the measurement results give $P_{z_1}=P_{+,E_1}+P_{-,E_1}$. (ii) Project measurement after exchanging the population of $|+, E_1\rangle$ and $|+, H_1\rangle$ by a $\pi$ pulse. The population stay in $|\pm E_1\rangle$ will be $P_{+,H_1}, P_{-,E_1}$ and the measurement results give $P_{z_2}=P_{+,H_1}+P_{-,E_1}$. (iii) Project measurement after exchanging the population in $|+, E_1\rangle$ and $|-, H_1\rangle$ by a $\pi$ pulse. The population stay in $|\pm E_1\rangle$ will be $P_{-,H_1}, P_{-,E_1}$ and the measurement results give $P_{z_3}=P_{-,E_1}+P_{-,H_1}$. Summarizing these results, we have,
\begin{eqnarray*}
P_{z_1}=P_{+,E_1}+P_{-,E_1},\\ P_{z_2}=P_{+,H_1}+P_{-,E_1},\\ P_{z_3}=P_{-,E_1}+P_{-,H_1}.
\end{eqnarray*}
Together with the normalization to 2 (that is, $P_{+,E_1}+P_{+,H_1}+P_{-,E_1}+P_{-,H_1}=2$), we can derive $P_{\pm,E_1 (H_1)}$ once $P_{z_1}, P_{z_2}, P_{z_3}$ are experimentally obtained. The $z$ components of Bloch vectors of $S_+(S_-)$ are connected to the population by $\langle\sigma_z^\tau\rangle'=2P_{\tau, E_1}-1$. The $y$ Bloch components $\langle\sigma_{y}^\tau\rangle'$ are achieved by similar measurement procedures as with $\langle\sigma_z^\tau\rangle'$, after an additional $\pi/2$ pulse $\chi_{y}(\frac{\pi}{2})$ rotates along $y$ axes. $\langle\sigma_{x}^\tau\rangle'$ components are also needed to accomplish the rotation (11), which are obtained by a similar operation as with $\langle\sigma_{y}^\tau\rangle'$.  To reduce measurement errors, a group of opposite rotations $\chi_{x, y}(-\frac{\pi}{2})$ are needed. The final results are determined by the difference between the results of $\chi_{x, y}(\pm\frac{\pi}{2})$. In the end, we can derive the three Bloch components $\langle\sigma_n^\tau\rangle'$ and then calculate $\langle\sigma_n^\tau\rangle$ by using Eq. (\ref{Rotation}).

In Fig. S4 and S5, the typical experimental results of $P_{\{x,y,z\}\{1,2,3\}}$ and $\langle\sigma^\pm_{x,y,z}\rangle'$  are shown under the conditions of $M=2B$ and $g=0/0.15A$ when driving the parameters in Fig. 2(a). Diamonds, squares, and circles are experimental data and the curves are theoretical curves. Based on the data in Fig. S4 and S5 and Eq. (\ref{Rotation}), the Bloch components $\langle\sigma_{x,y,z}^\pm\rangle$ of each pseudospin can be obtained, and the results are plotted in Fig. 2(c)(d) and Fig. S6(a)(b).  For the convenience of observation, the data in the trajectories in Fig.2(d) and Fig. S6(b) have been averaged  every 20 $\mu s$ for $T=500$$\mu s$ in each experimental cycle.